
\magnification=1200
\input macro1.tex                
\input epsf           

\font\bigbf=cmbx10 scaled\magstep1
\input macro2.tex                
\def\fig#1#2{\sspacing\hangindent=.75truein
\noindent \hbox to .75truein{Fig.\ #1.\hfil}#2
\hspacing\vskip 10pt}

\def\eqqqalignno#1{\tabskip=0em plus 50em\openup2\jot
\halign to \hsize{\hfil$
\displaystyle{##\null}$\tabskip=0pt &$
\displaystyle{##\qquad}$\hfil\tabskip=0pt &\hfil$
\displaystyle{##\null}$\tabskip=0pt &$\displaystyle{##}$\hfil\tabskip=1em
plus 50em &\hfil{##}\tabskip=0pt \crcr #1\crcr}}

\def\FR{F_{{\rm R}}}
\def\GR{G_{{\rm R}}}
\def\Gmult{\circ}
\def\calM#1{{\cal M}_{#1}}

\hspacing
%
%

\titletwolinesmod{DECORATED-BOX-DIAGRAM-CONTRIBUTIONS}
{TO BHABHA SCATTERING. (I)}

\author{G\"oran F\"aldt\footnote{${}^*$}{faldt@tsl.uu.se,
osland@vsfys1.fi.uib.no}}
\address{Gustaf Werners Institut, Box 535}
{S-751\ 21 Uppsala, Sweden}
\vskip 7 pt
\centerline{and}

\author{Per Osland${}^*$}
\address{Department of Physics, University of Bergen, All\'egt.~55}
{N-5007 Bergen, Norway}

\abstract{
We evaluate, in the high-energy limit, $s\gg|t|\gg m^2\gg\lambda^2$,
the sum of amplitudes corresponding to a class of Feynman diagrams
describing two-loop virtual photonic corrections to Bhabha scattering.
The diagrams considered are box and crossed box diagrams
with an extra photon decorating one of the fermion lines.
The mathematical method employed is that of Mellin transforms.
In the eikonal approximation, this sum of two-loop amplitudes has
previously been evaluated, and found to be equal to the sum of the
box and crossed box amplitudes, multiplied by the electric
form factor of the electron.
We obtain a similar factorization, but with the form factor
replaced by another expression involving
the logarithms $\log(\lambda^2/m^2)$ and $\log(\lambda^2/|t|)$.
}

\vfil
\eject
\newsec{Introduction}
In three earlier papers we initiated the study of virtual,
photonic corrections to Bhabha scattering
\ref{\refBFOone}{K.~S.\ Bj\o rkevoll, G.\ F\"aldt and P.\ Osland,
Nucl.\ Phys.\ {B386} (1992) 280},
\ref{\refBFOtwo}{K.~S.\ Bj\o rkevoll, G.\ F\"aldt and P.\ Osland,
Nucl.\ Phys.\ {B386} (1992) 303},
\ref{\refBFOthree}{K.~S.\ Bj\o rkevoll, G.\ F\"aldt and P.\ Osland,
Nucl.\ Phys.\ {B391} (1993) 591}.
The kinematical region considered, is that of small momentum transfers,
which is crucial for luminosity measurements at LEP,
and the diagrams considered are the six ladder-like diagrams.
In the present paper, we report results for another set of diagrams,
namely box diagrams with an extra photon decorating
one of the fermion lines.  There are eight diagrams in this set,
the sum of which is gauge invariant.
The diagrams are shown in fig.~1.

Neglecting terms linear in $\gamma_5$ in each current,
we write the Feynman amplitude for a single diagram as {\refBFOone}
$${\cal M}={i\alpha^3\over4\pi}\, F_{00}(s,t)
[\bar u(p_1')\gamma^\mu u(p_1)]\;
[\bar v(p_2)\gamma_\mu v(p_2')] .
\eqnref{\eqFtwoloop}$$
For a two-loop diagram the function $F_{00}(s,t)$ can be parametrized as
$$\eqalignno{
F_{00}(s,t)
&=\int_0^1\ldots\int_0^1 \d\alpha_1\ldots \d\alpha_7
\delta(1-\sum\alpha_i){1\over \Lambda(\alpha)^3} \cr
&\null\times\biggl[{2N_{III}\over D(\alpha)^3}
+{N_{II}\over 2D(\alpha)^2}
+{N_{I}\over 4D(\alpha)} \biggr], \eqalref{\eqFzerozero}}$$
with
$$D(\alpha)=D_ss+D_tt+D_mm^2+D_\lambda\lambda^2+\ieps, \eqnref{\eqDalpha}$$
but we shall also employ a parametrization based on the underlying box diagram.

We find that, asymptotically, for
$\lambda^2\ll m^2\ll|t|\ll s$,
and when we sum the contributions from the eight diagrams of fig.~1,
$$F_{00}^{(\rm sum)}(s,t)=-{4i\pi\over t}
\biggl[\log{\lambda^2\over m^2} +\log{\lambda^2\over|t|} +7 \biggr]
\log{\lambda^2\over|t|}. \eqno(\nr)$$
In this result we have discarded terms that vanish as $1/s$
for large values of $s$.
The eight diagrams where the photon decoration is attached to the lower line
give a summed contribution identical to the above one.

It should be noted that the amplitude of each individual Feynman diagram
contains terms that are proportional to both one and two powers of
$\log s$. In the sum over the set of eight diagrams, however, the terms
involving powers of $\log s$ cancel.
\newsec{Mathematical method}
The asymptotic values of the function $F_{00}(s,t)$ are calculated by
means of a Mellin transform
(see appendix~A).  The variable $s$ is considered to be very large
compared with $|t|$.  We know that $F_{00}(s,t)$,
at least the contribution from an individual diagram,
has powers of logarithms in $s$.
We define the Mellin transform of $F_{00}(s,t)$ through
$$\tilde I(\zeta,t)=\int_0^\infty \d s\, s^{-\zeta-1}\, F_{00}(s,t).
\eqno(\nr)$$
This choice is suitable for functions $F_{00}$ that behave asymptotically
as $1/t$, as opposed to $1/s$.
We expand $\tilde I(\zeta,t)$ around $\zeta=0$ and get
$$\tilde I(\zeta,t)={2A_3\over\zeta^3} +{A_2\over\zeta^2}
+{A_1\over\zeta} +\hbox{const.} \eqnref{\eqtwotwo} $$
This gives, when inverted, for large values of $s$
(see appendix~A):
$$F_{00}(s,t)=A_3\log^2(s)+A_2\log(s)+A_1. \eqno(\nr)$$
Thus, $A_1$ determines the constant, $A_2$ the linear logarithm, and
$A_3$ the quadratic logarithm.

Now, to every diagram there is a crossed (cr) diagram.
The amplitude for the crossed diagram is
$$\eqalignno{
F_{00}^{\hbox{\eightrm (cr)}}(s,t)&=-F_{00}(u,t) \cr
&=-A_3(\log(s)+i\pi)^2-A_2(\log(s)+i\pi)-A_1.
\eqalref{\eqcrossed}}$$
The addition of the amplitudes of a diagram and its crossed diagram yields
$$F_{00}(s,t)+F_{00}^{\hbox{\eightrm (cr)}}(s,t)
=-2i\pi A_3\log(s) -i\pi A_2 +\pi^2 A_3. \eqnref{\eqtvaafem}$$
Thus, the total amplitude will at most have linear logarithms in $s$.
In the sum {\eqtvaafem} the dependence on $A_1$ drops out.

As mentioned in the Introduction,
it turns out that
when we sum the contributions from all diagrams of fig.~1, then $A_3=0$.
Thus, the amplitude will not have any dependence on $\log s$.
\newsubsec{THEOREM~1}
If we exploit the Cheng-Wu theorem
\ref{\refCW}{H.\ Cheng and T.~T.\ Wu,
{\it Expanding Protons:  Scattering at High Energies},
(The MIT Press, Cambridge, Massachusetts, 1987)}\
(see also Appendix~A of {\refBFOtwo})
for a homogeneous integrand, and rescale one (or more)
of the variables that run to $\infty$, then we get
$$\eqalignno{
&\int_0^1\d\alpha_1\cdots\int_0^1\d\alpha_n\,
\delta(1-\sum_{i=1}^n\alpha_i)\, f(\alpha_1,\ldots,\alpha_n) \cr
&=\prod_{i=1}^m\biggl\{\kappa_i\int_0^\infty\d\beta_i\biggr\}
\prod_{j=m+1}^n\int_0^1\d\alpha_j\,
\delta(1-\sum_{k=m+1}^n\alpha_k) \cr
&\null\times
f(\kappa_1\beta_1,\ldots,\kappa_m\beta_m,\alpha_{m+1},\ldots,\alpha_n).
\eqalref{\eqTheoremone}}$$
We may now invert the Cheng-Wu transformation, and obtain
$$\eqalignno{
&\int_0^1\d\alpha_1\cdots\int_0^1\d\alpha_n\,
\delta(1-\sum_{i=1}^n\alpha_i)\, f(\alpha_1,\ldots,\alpha_n) \cr
&=\prod_{i=1}^m\kappa_i
\int_0^1\d\alpha_1\cdots\int_0^1\d\alpha_n\,
\delta(1-\sum_{i=1}^n\alpha_i)\,
f(\kappa_1\alpha_1,\ldots,\kappa_m\alpha_m,\alpha_{m+1},\ldots,\alpha_n). \cr
&\eqalref{\eqTheoremonebis}}$$
\newsubsec{THEOREM~2}
Integrations over Feynman parameters $\alpha_i$ of integrands that are
singular as $\alpha_i\to0$, are sometimes done by applying the theorem
$$\eqalignno{
F_n(\zeta)&=\int_0^\infty\cdots\int_0^\infty
\d\alpha_1\, \cdots \d\alpha_n\,(\alpha_1\cdots \alpha_n)^{\zeta-1}\,
f(\alpha_1\, \cdots,\alpha_n) \cr
&={1\over\zeta^n}\, f(0,\cdots,0)
+{1\over\zeta^{n-1}}\, \sum_{k=1}^n
\biggl\{\int_0^1\d\alpha_k\, {1\over\alpha_k}[f_k(\alpha_k)-f_k(0)] \cr
&+\int_1^\infty\d\alpha_k\, {1\over\alpha_k}\, f_k(\alpha_k) \biggr\}
+\Order(\zeta^{-n+2}) \cr
&={1\over\zeta^{n-1}}\sum_{k=1}^n
\int_0^\infty\d\alpha_k\, \alpha_k^{\zeta-1}\, f_k(\alpha_k)
-{n-1\over\zeta^n}\, f(0,\cdots,0) +\Order(\zeta^{-n+2}).
\eqalref{\eqTheoremtwo}}$$
Both forms are useful.  Here,
$$f_k(\alpha_k)=f(\alpha_k,\alpha_l=0, l\ne k). \eqno(\nr)$$
In particular,
$$f_k(0)=f(0,\cdots,0), \qquad k=1,\ldots,n. \eqno(\nr)$$

\newsec{Preliminaries for one- and two-loop diagrams}
Before discussing the details of the two-loop diagrams, it is useful to
collect a few general formulas for the one-loop and two-loop diagrams.
Such results are presented in the present section.
\newsubsec{GENERAL EVALUATION OF THE ONE-LOOP DIAGRAM}
Consider first the scalar one-loop diagram,
$$I=\int{\d^4q\over(2\pi)^4}\,
\prod_{i=1}^4
{1\over (Q_i+u_iq)^2-m_i^2+\ieps}, \eqno(\nr)$$
where the $Q_i$ are constant four-momenta and
the $u_i$ are constants $(=\pm1,0)$.
By Feynman parametrization the product can be written
$$\prod_{i=1}^4{1\over a_i}
=\Gamma(4)\int_0^1\ldots\int_0^1 \d\alpha_1\ldots \d\alpha_4
{\delta(1-\sum\alpha_i)\over
[\sum\alpha_i a_i +\ieps]^4}. \eqno(\nr)$$

We define the vector multiplication $\Gmult$ such that
the Feynman parameters $\alpha_i$ play the role of a diagonal metric, i.~e.,
$$\eqalignno{
u\Gmult u &=\sum \alpha_i u_i^2, \cr
u\Gmult Q &=\sum \alpha_i u_i Q_i, \cr
m\Gmult m &=\sum \alpha_i m_i m_i. \eqalref{\eqCthree} }$$
Then we can write
$$\eqalignno{
\tilde D&=\sum \alpha_i a_i \cr
&=(u\Gmult u) q^2 +2q\cdot(u\Gmult Q)
+(Q\Gmult Q) -(m\Gmult m). &(\nr)}$$

The integration over $q$ is straightforward,
$$\eqalignno{
I&={i\over(4\pi)^2}\int_0^1\ldots\int_0^1 \d\alpha_1\ldots \d\alpha_4 \,
\delta(1-\sum\alpha_i)\, {1\over[D+\ieps]^2}, &(\nr) \cr
\Lambda&=(u\Gmult u), &(\nr) \cr
D&=\Lambda[(Q\Gmult Q)-(m\Gmult m)] -(u\Gmult Q)^2. &(\nr)}$$

Consider next
$$\eqalignno{
M^{\mu\nu}
&=\int{\d^4q\over(2\pi)^4} \;
{l_1^\mu\over l_1^2-m_1^2+\ieps} \,
{l_2^\nu\over l_2^2-m_2^2+\ieps} \cr
&\null\times
{1\over l_3^2-m_3^2+\ieps} \,{1\over l_4^2-m_4^2+\ieps}, &(\nr) }$$
with
$$l_i=Q_i+u_i q. \eqnref{\eqlsubi}$$
Using the methods outlined in Appendix~A of
ref.~{\refBFOone},
one finds
$$\eqalignno{
M^{\mu\nu}
&={i\over16\pi^2}\, \int_0^1\ldots\int_0^1 \d\alpha_1\ldots \d\alpha_4 \,
\delta(1-\sum\alpha_i)\, {1\over \Lambda^2} \cr
&\null\times
\biggl[
{1\over D^2} k_1^\mu k_2^\nu
+{1\over 2D}\, a_{12}g^{\mu\nu} \biggr], &(\nr)}$$
with $\Lambda$ and $D$ as above, and
$$\eqalignno{
k_i&=\Lambda Q_i
-(u\Gmult Q)u_i,  &(\nr) \cr
a_{12}&=u_1 u_2. &(\nr)}$$

We shall also need the integral
$$M^\mu=\int{\d^4q\over(2\pi)^4} \;
{l_1^\mu\over l_1^2-m_1^2+\ieps} \,
\prod_{i=2}^4 {1\over l_i^2-m_i^2+\ieps}, \eqno(\nr)$$
with $l_i$ as in eq.~{\eqlsubi}. In this case we have
$$M^\mu={i\over16\pi^2}\, \int_0^1\ldots\int_0^1 \d\alpha_1\ldots \d\alpha_4 \,
\delta(1-\sum\alpha_i)\, {1\over \Lambda}\, {k_1^\mu\over D^2}.
\eqno(\nr)$$

The methods presented in this section are used in the evaluation of the
vertex diagram in appendix~B, as well as in the evaluation of diagram (b)
in sect.~5 and in the evaluation of diagram (d) in sect.~6.
\newsubsec{THE BOX DIAGRAM}
We need some results for the box diagram, fig.~2, for the case when the masses
associated with the internal fermion lines ($m_1$ and $m_2$)
are different from those of the external ones ($m$).
Ignoring $\gamma_5$-parts, we have
$${\cal M}=i\alpha^2\, G_{00}(s,t)
[\bar u(p_1')\gamma^\mu u(p_1)]\;
[\bar v(p_2)\gamma_\mu v(p_2')],
\eqnref{\eqGbox}$$
$$\eqalignno{
G_{00}(s,t)
&=\int_0^1\ldots\int_0^1 \d z_1 \d z_2 \d z_3 \d z_4
\delta(1-\sum z_i){1\over \Lambda(z)^2} \cr
&\null\times\biggl[{N_{II}\over D(z)^2}
+{N_{I}\over 2D(z)} \biggr]. \eqalref{\eqGzerozero}}$$
Here,
$$D(z)=D_{\hbox{\eightrm box}}(z)
-z_1\Lambda(z)(m_1^2-m^2)-z_2\Lambda(z)(m_2^2-m^2), \eqno(\nr)$$
where
$$D_{\hbox{\eightrm box}}(z)=D_ss+D_tt+D_mm^2
+D_\lambda\lambda^2+\ieps, \eqnref{\eqDbox}$$
is the ordinary box-diagram denominator, with
$$\eqalignno{
D_s&=z_1z_2, &(\nr) \cr
D_t&=z_3z_4, &(\nr) \cr
-D_m&=(z_1+z_2)^2, &(\nr) \cr
-D_\lambda&=(z_3+z_4)\Lambda(z), &(\nr)\cr
\Lambda(z)&=z_1+z_2+z_3+z_4, &(\nr)}$$
and $m_1$ and $m_2$ the masses associated with the upper and lower internal
fermion lines of fig.~2.
The numerators in eq.~{\eqGzerozero} are
$$\eqalignno{
N_{II}&=-s[2(z_1+z_3+z_4)
(z_2+z_3+z_4)+z_1z_2], \eqalref{\eqNIIbox} \cr
N_{I}&=10. &(\nr)}$$
These results will be used in sects.~5 and 6.

In our evaluations of one- and two-loop amplitudes
we often encounter the integrals
$$\eqalignno{
C_{20}(\lambda^2)&=\int_0^1 \d x\,
{1\over x(1-x)+\lambda^2} \cr
&=-2\log\lambda^2 +\Order(\lambda^2), \eqalref{\eqCtwozero} \cr
C_{21}(\lambda^2)&=\int_0^1 \d x\,
{1\over x(1-x)+\lambda^2} \,
\log[x(1-x)+\lambda^2] \cr
&=-\log^2\lambda^2-{\pi^2\over3} +\Order(\lambda^2). \eqalref{\eqCtwoone} }$$

For Bhabha scattering, the uncrossed box diagram gives,
in the asymptotic limit we are considering,
$$\eqalignno{
G_{00}(s,t)&=-{2\over t}\,\log\biggl({s\over |t|}\biggr)
C_{20}\biggl({\lambda^2\over|t|}\biggr) \cr
&={4\over t}
\log\biggl({s\over |t|}\biggr)
\log\biggl({\lambda^2\over|t|}\biggr).
\eqalref{\eqGboxBhabha} }$$

The invariant function for the crossed box diagram (cr) is given by
$$G_{00}^{\hbox{\eightrm (cr)}}(s,t)=-G_{00}(u,t). \eqno(\nr)$$
\newsubsec{GENERAL EVALUATION OF THE TWO-LOOP DIAGRAM}
Consider first the scalar two-loop diagram,
$$I=\int{\d^4q_1\over(2\pi)^4}\, \int{\d^4q_2\over(2\pi)^4}
\prod_{i=1}^7
{1\over (Q_i+u_iq_1+v_iq_2)^2-m_i^2+\ieps}, \eqno(\nr)$$
where the $Q_i$ are constant four-momenta and
the $u_i$ and $v_i$ are constants $(=\pm1,0)$.
By Feynman parametrization the product can be written
$$\prod_{i=1}^7{1\over a_i}
=\Gamma(7)\int_0^1\ldots\int_0^1 \d\alpha_1\ldots \d\alpha_7
{\delta(1-\sum\alpha_i)\over
[\sum\alpha_i a_i +\ieps]^7}. \eqno(\nr)$$

We define the vector multiplication $\Gmult$ as in {\eqCthree}.
Then
$$\eqalignno{
\tilde D&=\sum \alpha_i a_i \cr
&=(u\Gmult u) q_1^2 +2q_1\cdot[(u\Gmult Q) +(u\Gmult v) q_2] \cr
&+(v\Gmult v) q_2^2 +2q_2\cdot(v\Gmult Q) +(Q\Gmult Q) -(m\Gmult m). &(\nr)}$$

The integrations over $q_1$ and $q_2$ are straightforward,
$$\eqalignno{
I&={-2\over(4\pi)^4}\int_0^1\ldots\int_0^1 \d\alpha_1\ldots \d\alpha_7 \,
\delta(1-\sum\alpha_i)\, {\Lambda\over[D+\ieps]^3}, &(\nr) \cr
\Lambda&=(u\Gmult u)(v\Gmult v)-(u\Gmult v)^2, \eqalref{\eqLambdagen} \cr
D&=\Lambda[(Q\Gmult Q)-(m\Gmult m)] \cr
&-(u\Gmult u)(v\Gmult Q)^2 -(v\Gmult v)(u\Gmult Q)^2
+2(u\Gmult v)(u\Gmult Q)\cdot(v\Gmult Q). \eqalref{\eqDgen} }$$

Consider next
$$\eqalignno{
M^{\mu\nu\rho\sigma}
&=\int{\d^4q_1\over(2\pi)^4}\, \int{\d^4q_2\over(2\pi)^4} \cr
&\null\times
{l_1^\mu\over l_1^2-m_1^2+\ieps} \,
{l_2^\nu\over l_2^2-m_2^2+\ieps} \,
{l_3^\rho\over l_3^2-m_3^2+\ieps} \,
{l_4^\sigma\over l_4^2-m_4^2+\ieps} \cr
&\null\times
{1\over l_5^2-m_5^2+\ieps} \,{1\over l_6^2-m_6^2+\ieps} \,
{1\over l_7^2-m_7^2+\ieps}, &(\nr) }$$
where
$$l_i=Q_i+u_i q_1 +v_i q_2. \eqno(\nr)$$
Using the methods outlined in Appendix~A of
{\refBFOone},
one finds
$$\eqalignno{
M^{\mu\nu\rho\sigma}
&=-{1\over(16\pi^2)^2}\, \int_0^1\ldots\int_0^1 \d\alpha_1\ldots \d\alpha_7 \,
\delta(1-\sum\alpha_i)\, {1\over \Lambda^3} \cr
&\null\times
\biggl\{
{2\over D^3} k_1^\mu k_2^\nu k_3^\rho k_4^\sigma \cr
&+{1\over 2D^2}\biggl[
a_{12}g^{\mu\nu}k_3^\rho k_4^\sigma +a_{13}g^{\mu\rho}k_2^\nu k_4^\sigma
+a_{14}g^{\mu\sigma}k_2^\nu k_3^\rho \cr
&\phantom{{1\over 2D^2}\biggl[}
+a_{23}g^{\nu\rho}k_1^\mu k_4^\sigma
+a_{24}g^{\nu\sigma}k_1^\mu k_3^\rho +a_{34}g^{\rho\sigma}k_1^\mu k_2^\nu
\biggr] \cr
&+{1\over 4D}\biggl[
 a_{12}a_{34}g^{\mu\nu}g^{\rho\sigma}
+a_{13}a_{24}g^{\mu\rho}g^{\nu\sigma}
+a_{14}a_{23}g^{\mu\sigma}g^{\nu\rho}\biggr] \biggr\}, &(\nr)}$$
with $\Lambda$ and $D$ as above, eqs.~{\eqLambdagen} and {\eqDgen}, and
$$\eqalignno{
k_i&=\Lambda Q_i
-(u\Gmult u)(v\Gmult Q)v_i
-(v\Gmult v)(u\Gmult Q)u_i \cr
&+(u\Gmult v)[(u\Gmult Q)v_i+(v\Gmult Q)u_i], &(\nr) \cr
a_{ij}&=(u\Gmult u)v_i v_j +(v\Gmult v)u_i u_j
-(u\Gmult v)(u_i v_j +v_i u_j)
-\delta_{ij}{\Lambda\over\alpha_i}. &(\nr)}$$
\newsec{Diagram (a)}
The Feynman amplitude for diagram (a) of fig.~1 is given by
$$\eqalignno{
\calM{(a)}=&\int{\d^4q_1\over(2\pi)^4} \int{\d^4q_2\over(2\pi)^4} \;
\bar u(p_1')(-ie\gamma_\gamma)
{i(\rlap/ p_1'-\rlap/ q_2 +m) \over (p_1'-q_2)^2-m^2+i\epsilon} \cr
\times&(-ie\gamma_\beta)
{i(\rlap/ p_1-\rlap/ q_1-\rlap/ q_2 +m) \over (p_1-q_1-q_2)^2-m^2+i\epsilon}
(-ie\gamma_\alpha) \cr
\times&{i(\rlap/ p_1-\rlap/ q_2 +m) \over (p_1-q_2)^2-m^2+i\epsilon}
(-ie\gamma^\gamma) u(p_1) \cr
\times&\bar v(p_2)(-ie\gamma^\alpha)
{i(-\rlap/ p_2-\rlap/ q_1 +m) \over (p_2+q_1)^2-m^2+i\epsilon}
(-ie\gamma^\beta) v(p_2') \cr
\times&{-i\over q_1^2-\lambda^2+i\epsilon}\;
{-i\over q_2^2-\lambda^2+i\epsilon}\;
{-i\over (p_1-p_1'-q_1)^2-\lambda^2+i\epsilon}. &(\nr)}$$
By the general method of sect.~3.3, we obtain after integration over momenta
an expression of the form
{\eqFtwoloop}--{\eqDalpha}, with
$$\eqalignno{
D(\alpha)&=D_ss+D_tt+D_mm^2+D_\lambda\lambda^2 +\ieps \cr
&\equiv D_ss+\calD(\alpha), \eqalref{\eqcalD} \cr
D_s&=\alpha_2\alpha_4\alpha_5, \eqalref{\eqDs} \cr
D_t&=\alpha_6\alpha_7(\alpha_1+\alpha_2+\alpha_3+\alpha_5)
+\alpha_1\alpha_6(\alpha_2+\alpha_3)+\alpha_3\alpha_7(\alpha_1+\alpha_2) \cr
&+\alpha_1\alpha_3(\alpha_2+\alpha_4), \eqalref{\eqDt}\cr
-D_m&=(\alpha_1+\alpha_2+\alpha_3)
[(\alpha_1+\alpha_2+\alpha_3)(\alpha_2+\alpha_4+\alpha_6+\alpha_7)
+\alpha_4^2-\alpha_2^2] \cr
&+\alpha_5(\alpha_2+\alpha_4)^2, &(\nr)\cr
-D_\lambda&=(\alpha_5+\alpha_6+\alpha_7)\Lambda(\alpha), &(\nr)\cr
\Lambda(\alpha)&=(\alpha_1+\alpha_2+\alpha_3+\alpha_5)
(\alpha_2+\alpha_4+\alpha_6+\alpha_7)-\alpha_2^2, \eqalref{\eqLambda}}$$
and the numerator functions
$$\eqalignno{
N_{III}&=2s^2\alpha_2^3\alpha_4^2\alpha_5^2(\alpha_4+\alpha_6+\alpha_7),
\eqalref{\eqNthree}\cr
N_{II}&=-4s\alpha_5[2(\alpha_2+\alpha_6+\alpha_7)
(\alpha_4+\alpha_6+\alpha_7)\Lambda(\alpha)
+\alpha_2\alpha_4\Lambda(\alpha) \cr
&+3\alpha_2^2\alpha_4(\alpha_4+\alpha_6+\alpha_7)],
\eqalref{\eqNtwo}\cr
N_{I}&=8[5\Lambda(\alpha)-3\alpha_2(\alpha_4+\alpha_6+\alpha_7)].
\eqalref{\eqNone}}$$
The $\alpha$'s are associated with the internal lines of the
Feynman diagram as shown in fig.~3.

We now calculate $F_{00}^{\rm (a)}(s,t)$.
A Mellin transformation shows that the contributions from the
$N_{III}$ and $N_{I}$ terms of eq.~{\eqFzerozero} go as $1/s$
and may be neglected.
This is demonstrated in appendix~C.
We are thus left with [cf.\ eq.~{\eqFzerozero}]
$$F_{00}^{\rm(a)}(s,t)
=\int_0^1\ldots\int_0^1 \d\alpha_1\ldots \d\alpha_7
\delta(1-\sum\alpha_i){1\over \Lambda(\alpha)^3} \,
{N_{II}\over 2D(\alpha)^2},
\eqnref{\eqFazerozero}$$
where we put $|t|=1$, and perform a Mellin transformation,
$$\eqalignno{
\tilde I^{\rm(a)}(\zeta)
&=\int_0^\infty \d s\,s^{-\zeta-1}\; F_{00}^{\rm(a)}(s,t) \cr
&={\Gamma(1-\zeta)\Gamma(1+\zeta)\over\Gamma(2)}\,
\int_0^1\ldots\int_0^1 \d\alpha_1\ldots \d\alpha_7\,
\delta(1-\sum_{j=1}^7\alpha_j) \cr
&\null\times
{N_{II}/(2s) \over \Lambda(\alpha)^3} \,
(\alpha_2\alpha_4\alpha_5)^{\zeta-1}\, \calD(\alpha)^{-1-\zeta},
\eqalref{\eqMellin} \cr
\calD(\alpha)&=-D_t+D_mm^2+D_\lambda\lambda^2+\ieps. \eqalref{\eqdlower}}$$

The dominant singularity of $\tilde I^{\rm(a)}(\zeta)$
is obviously at $\alpha_2\simeq\alpha_4\simeq0$.
There is no singularity associated with $\alpha_5\simeq0$, due to the
factor $\alpha_5$ in $N_{II}$.
Nevertheless, $\tilde I^{\rm(a)}(\zeta)$ is in reality of order $\zeta^{-3}$.
In order to clarify the location of the most singular contribution we make
a transformation of the variables.
First, we apply the Cheng-Wu theorem in the variables
$\alpha_2$, $\alpha_4$, $\alpha_6$, and $\alpha_7$.
Then, we introduce the following new integration variables,
$$\alpha_4+\alpha_6+\alpha_7={1\over\rho}, \qquad 0\le\rho\le\infty, $$
$$\alpha_4={1\over\rho}\,(1-x),
\qquad \alpha_6={1\over\rho}\,xz,
\qquad \alpha_7={1\over\rho}\,x(1-z), \eqno(\nr)$$
and get
$$\eqalignno{
\tilde I^{\rm(a)}(\zeta)
&=-e^{-i\pi\zeta}\,
\int_0^1 \int_0^1 \int_0^1 \d\alpha_1 \d\alpha_3 \d\alpha_5 \,
\delta(1-\alpha_1-\alpha_3-\alpha_5)\,
\int_0^\infty \d\alpha_2 \int_0^\infty \d\rho \cr
&\times \int_0^1 x\,\d x \int_0^1 \d z
[\alpha_2\rho(1-x)]^{\zeta-1}\,
\alpha_5^{\zeta-1}{N_{II}/(2s)\over \Lambda(\alpha)^3}
(-\rho^2\calD)^{-1-\zeta}. &(\nr)}$$
This shows that the leading singularity is at $\alpha_2\simeq0$,
$\rho\simeq0$, and $x\simeq1$, provided it is not softened by any
of the remaining factors in the integrand.

In terms of the new variables, we get from eq.~{\eqdlower}
and eqs.~{\eqDt}--{\eqLambda},
$$-\rho^2\calD=a+b\rho+c\rho^2, \eqno(\nr)$$
with
$$\eqalignno{
a&=(1+\alpha_2)\bigl[x^2z(1-z) +(1-x)^2m^2 +x\lambda^2\bigr], \cr
b&=\alpha_1\alpha_3 +\alpha_2x(\alpha_1z+\alpha_3(1-z)) \cr
&+\bigl[(\alpha_1+\alpha_2+\alpha_3)^2 +2(1-x)\alpha_2\alpha_5\bigr]m^2
+\bigl[\alpha_5(1+\alpha_2)+x\alpha_2\bigr]\lambda^2, \cr
c&=\alpha_2\bigl\{\alpha_1\alpha_3 +[(\alpha_1+\alpha_3)^2 +\alpha_2]m^2
+\alpha_5\lambda^2\bigr\}, \eqalref{\eqabc}}$$
where we have made use of the identity $\alpha_1+\alpha_3+\alpha_5=1.$
{}From equations {\eqLambda} and {\eqNtwo} we get
$$\eqalignno{
\rho\Lambda(\alpha)&=\rho\alpha_2+(1+\alpha_2), \cr
\rho^3N_{II}/(2s)&=-2\alpha_5\bigl[
2x(\rho\Lambda(\alpha))
+\rho\alpha_2(3-x)(\rho\Lambda(\alpha))
+3\rho\alpha_2^2(1-x) \bigr]. &(\nr)}$$

We can now confirm our previous presumption that the leading contribution comes
from $\alpha_2\simeq0$, $\rho\simeq0$, and $x\simeq1$.
We are interested in the leading and subleading contributions.
Therefore, at least two of these conditions must be fulfilled.
In particular, this means that we can put $c\rho=0$ and
$\rho\Lambda(\alpha)=1+\alpha_2$ in the integrand.
When this is done we get a simpler expression
$$\eqalignno{
\tilde I^{\rm(a)}(\zeta)
&=4e^{-i\pi\zeta}
\int_0^1 \int_0^1 \int_0^1 \d\alpha_1 \d\alpha_3 \d\alpha_5
\,\delta(1-\alpha_1-\alpha_3-\alpha_5)\,
\alpha_5^\zeta\int_0^1\d z \cr
&\times \int_0^\infty \d\alpha_2 \int_0^\infty \d\rho \int_0^1 \d x
[\alpha_2\rho(1-x)]^{\zeta-1}\,
{x^2\over (1+\alpha_2)^2}\, {1\over (a+b\rho)^{1+\zeta}}. &(\nr)}$$
We can immediately do the integration over $\rho$, using the formula
$$\eqalignno{
\int_0^\infty \d\rho \, \rho^{\zeta-1} (a+b\rho)^{-\zeta-1}
&=b^{-\zeta} a^{-1} \int_0^\infty \d x\, x^{\zeta-1}\, (1+x)^{-1-\zeta} \cr
&=b^{-\zeta} a^{-1} B(\zeta,1)={1\over\zeta}\, b^{-\zeta} a^{-1},
\eqalref{\eqIntbeta}}$$
where $B$ is the Euler $\beta$-function,
$$\int_0^\infty\d x\, x^{\mu-1}(1+x)^{-\nu}
=B(\mu,\nu-\mu)
={\Gamma(\mu)\Gamma(\nu-\mu)\over\Gamma(\nu)}. \eqnref{\eqBeta}$$

We get
$$\eqalignno{
\tilde I^{\rm(a)}(\zeta)
&={4\over\zeta}\,e^{-i\pi\zeta}
\int_0^1 \int_0^1 \int_0^1 \d\alpha_1 \d\alpha_3 \d\alpha_5
\,\delta(1-\alpha_1-\alpha_3-\alpha_5)\,
\alpha_5^\zeta\int_0^1\d z \cr
&\times \int_0^\infty \d\alpha_2 \int_0^1 \d x
[\alpha_2(1-x)]^{\zeta-1}\,
{x^2\over (1+\alpha_2)^2}\, {1\over a}\, b^{-\zeta}. \eqalref{\eqIatilde}}$$
We now apply our singularity theorem {\eqTheoremtwo} and write
$$\tilde I^{\rm(a)}(\zeta)=\tilde I_0(\zeta) +\tilde I_1(\zeta)
+\tilde I_2(\zeta). \eqnref{\eqIdecomp}$$
The first term contains the leading singularity,
$$\tilde I_0(\zeta)={2\over\zeta^3}\, e^{-i\pi\zeta}\,
f_0\, C_{20}(\lambda^2). \eqno(\nr)$$
This result is obtained by putting $\alpha_2=0$ and $x=1$ in
the integrand of {\eqIatilde}.
We have
$$\eqalignno{
f_0&=2\int_0^1 \int_0^1 \int_0^1 \d\alpha_1 \d\alpha_3 \d\alpha_5
\,\delta(1-\alpha_1-\alpha_3-\alpha_5)\,
[1-\zeta\log(\alpha_1\alpha_3)+\zeta\log(\alpha_5)] \cr
&=1+{\textstyle 3\over2}\zeta. \eqalref{\eqfzero}}$$
The function $C_{20}(\lambda^2)$ is given by eq.~{\eqCtwozero}.

We then calculate $\tilde I_1(\zeta)$ of eq.~{\eqIdecomp}.
It is the subleading contribution when $x=1$,
and given by
$$\tilde I_1(\zeta)={2\over\zeta^2}\, f_1\, C_{20}(\lambda^2), \eqno(\nr)$$
with
$$f_1=\int_0^1\d\alpha\, {1\over\alpha}
\biggl[{1\over(1+\alpha)^3}-1\biggr]
+\int_1^\infty{\d\alpha\over\alpha}\, {1\over(1+\alpha)^3} =-{3\over2}.
\eqno(\nr)$$

Then we finally calculate $\tilde I_2(\zeta)$, the subleading contribution
when $\alpha_2=0$,
$$\eqalignno{
\tilde I_2(\zeta)
&={4\over\zeta^2}
\int_0^1 \int_0^1 \int_0^1 \d\alpha_1 \d\alpha_3 \d\alpha_5
\,\delta(1-\alpha_1-\alpha_3-\alpha_5) \cr
&\null\times\int_0^1 \d z \int_0^1 \d x\,
{1\over 1-x}\biggl[{x^2\over a_0}-{1\over a_{00}}\biggr],
\eqalref{\eqItwotilde}}$$
with [cf.~eq.~{\eqabc}]
$$\eqalignno{
a_0&=a(\alpha_2=0)
=x^2z(1-z)+(1-x)^2m^2+x\lambda^2, \cr
a_{00}&=a(\alpha_2=0,x=1). \eqalref{\eqaa} }$$

The integral over the $\alpha$-variables is done in eq.~{\eqfzero}.
The remaining two integrations are straightforward but lengthy.
We are led to an integral which appears frequently in our calculations,
$$\eqalignno{
R(m^2,\lambda^2)
&\equiv\int_0^1\d \alpha\int_0^1{\d\beta\over\beta}
\biggl[{1\over 1+\beta}\,{1\over N} -{1\over N_0}\biggr] \cr
&+\int_0^1\d \alpha\int_1^\infty {\d \beta\over \beta}\, {1\over 1+\beta}\,
{1\over N}, \eqalref{\eqRdef}}$$
with
$$\eqalignno{
N&=\alpha(1-\alpha)+\beta^2 m^2+(1+\beta)\lambda^2, \cr
N_0&=N(\beta=0)=\alpha(1-\alpha)+\lambda^2. &(\nr)}$$
It is calculated in appendix~D, where it is also shown that the present
integral, eq.~{\eqItwotilde}, over $z$ and $x$ is actually given by $R$.
Thus
$$\eqalignno{
\tilde I_2(\zeta)
&={2\over\zeta^2}\, R(m^2,\lambda^2) \cr
&=-{2\over\zeta^2}
\biggl[{1\over2}\log^2\biggl({\lambda^2\over m^2}\biggr)
+{5\pi^2\over6}\biggr]. &(\nr)}$$

We are now in a position to give the result for
$\tilde I^{\rm(a)}(\zeta)$ of eq.~{\eqIdecomp}.
To the order considered, and for $|t|=1$,
$$\tilde I^{\rm(a)}(\zeta)
={2\over\zeta^3}\, e^{-i\pi\zeta}\, C_{20}(\lambda^2)
+{2\over\zeta^2}\, R(m^2,\lambda^2).
\eqno(\nr)$$

\newsec{Diagrams (b) and (c)}
This section treats the contributions from
the diagrams (b) and (c) of fig.~1.
They can be considered as box diagrams with
vertex corrections.  We shall make use of the results for the
half-off-shell vertex function discussed in appendix~B, and
the results for the general (with unequal masses) box diagram
of section~3.

The amplitudes for the (b) and (c) diagrams are identical,
$$F_{00}^{\rm(c)}(s,t)=F_{00}^{\rm(b)}(s,t). \eqno(\nr)$$
Thus, it is sufficient to calculate $F_{00}^{\rm(b)}(s,t)$.

\newsubsec{DECOMPOSITION OF $F_{00}^{\rm(b)}(s,t)$}
We can write the amplitude $\calM{(b)}$ in terms of the renormalized
one-loop vertex correction $\Gamma^\alpha(p',p;q)$ as [see fig.~4]:
$$\eqalignno{
\calM{(b)}=&\int{\d^4q\over(2\pi)^4}
\bar u(p_1')(-ie\gamma^\beta)
{i(\rlap/ p_1-\rlap/ q +m) \over (p_1-q)^2-m^2+i\epsilon} \cr
\times&[-ie\Gamma^\alpha(p_1-q,p_1;q)]u(p_1) \cr
\times&\bar v(p_2)(-ie\gamma_\alpha)
{i(-\rlap/ p_2-\rlap/ q +m) \over (p_2+q)^2-m^2+i\epsilon}
(-ie\gamma_\beta) v(p_2') \cr
\times&{-i\over q^2-\lambda^2+i\epsilon}\;
{-i\over (p_1-p_1'-q)^2-\lambda^2+i\epsilon}. &(\nr)}$$

We start with the decomposition of the half-off-shell
vertex functions with $p=p_1$, $p'=p_1-q$, and $p_1^2=m^2$, eq.~(B.19),
$$\eqalignno{
\Gamma^\alpha(p_1-q,p_1;q)
&={\alpha\over2\pi}\int_0^1 \rho \d\rho \int_0^1 \d x \,
{1\over r_1[(p_1-q)^2-m^2] +r_2(q^2-\lambda^2)-M^2+\ieps} \cr
&\times\biggl\{
\gamma^\alpha \{n_1[(p_1-q)^2-m^2] +n_2(q^2-\lambda^2)\} \cr
&-2(1-\rho x)\,(\pslash_1-\qslash-m)[\rho x\,q^\alpha
+(1-\rho)p_1^\alpha]\biggr\}.  \eqalref{\eqGamma}}$$
Terms explicitly proportional to $m^2$ and $\lambda^2$ have been neglected.
Furthermore,
$$\eqqqalignno{
r_1&=\rho(1-\rho)x, & r_2&=\rho^2x(1-x), \cr
n_1&=(1-\rho)^2,    & n_2&=-(1-\rho)-\rho^2(1-x)^2. \eqalref{\eqmdef}}$$
and
$$M^2=\rho^2m^2 +(1-\rho)\lambda^2-r_2\lambda^2. \eqnref{\eqMtwo} $$

The formulas for the box diagram, in section~3.2, allow us to
take into account the vertex corrections
by suitable substitutions for the propagators in the box diagram,
$$F_{00}^{\rm(b)}(s,t)
=F_{00}^{\rm(b1)}(s,t)+F_{00}^{\rm(b2)}(s,t)
+F_{00}^{\rm(b3)}(s,t)+F_{00}^{\rm(b4)}(s,t), \eqnref{\eqFonethree}$$
with $F_{00}^{\rm(b1)}(s,t)$ and $F_{00}^{\rm(b2)}(s,t)$ corresponding
to the terms with the $n_1$ and $n_2$ factors in eq.~{\eqGamma}.
Similarly, $F_{00}^{\rm(b3)}(s,t)$ and $F_{00}^{\rm(b4)}(s,t)$
are the third and fourth terms.
\newsubsec{THE FUNCTION $F_{00}^{\rm(b1)}(s,t)$}
For the first term, we get after integration over the four-momentum $q$,
$$\eqalignno{
F_{00}^{\rm(b1)}(s,t)&=2\int_0^1 \rho \d\rho \int_0^1 \d x
\int_0^1\ldots\int_0^1 \d\alpha_1\ldots \d\alpha_4 \,
\delta(1-\sum\alpha_i){n_1\over \Lambda(\beta)^2} \cr
&\null\times\biggl[{N_{II}^{\hbox{\eightrm box}}(\beta)\over D(\beta)^2}
+{N_{I}^{\hbox{\eightrm box}}(\beta)\over 2D(\beta)} \biggr],
\eqalref{\eqFzerozeroone} }$$
with
$N_{II}^{\hbox{\eightrm box}}(\beta)$ and
$N_{I}^{\hbox{\eightrm box}}(\beta)$ given in section~3.2.
In particular,
$$N_{II}^{\hbox{\eightrm box}}(\beta)
=-s[2(\beta_1+\beta_3+\beta_4)(\beta_2+\beta_3+\beta_4)+\beta_1\beta_2].
\eqno(\nr)$$
Furthermore, the transformed Feynman parameters are given as
$$\eqqqalignno{
\beta_1&=\alpha_1 r_1, & \beta_2&=\alpha_2, \cr
\beta_3&=\alpha_3+\alpha_1 r_2,     & \beta_4&=\alpha_4, &(\nr)}$$
and
$$\eqalignno{
\Lambda(\beta)&=\beta_1+\beta_2+\beta_3+\beta_4 \cr
&=\alpha_1(r_1+r_2)+\alpha_2+\alpha_3+\alpha_4, \eqalref{\eqLambdab}}$$
$$\eqalignno{
D(\beta)&=D_{\hbox{\eightrm box}}(\beta)
-\alpha_1\Lambda(\beta) M^2 \cr
&=\beta_1\beta_2s+\beta_3\beta_4t
-[(\beta_1+\beta_2)^2+\alpha_1\Lambda(\beta)\rho^2]m^2 \cr
&\quad-[(\beta_3+\beta_4)+\alpha_1(1-\rho-r_2)]\Lambda(\beta)\lambda^2
+\ieps \cr
&\equiv\beta_1\beta_2s +d,  \eqalref{\eqfourseven}}$$
where $D_{\hbox{\eightrm box}}$,
eq.~{\eqDbox}, is the basic $D$ function for the
box diagram when the masses of the internal fermion lines are equal
to the external ones.

The evaluation of the $N_{I}^{\hbox{\eightrm box}}$-term
in eq.~{\eqFzerozeroone} is straightforward.
It gives a contribution that for large values of $s$
goes as $1/s$, with the leading term being proportional to $\log^3(s)/s$,
and is therefore neglected.
Putting $|t|=1$ and performing a Mellin transform, we get for the
$N_{II}^{\hbox{\eightrm box}}$-term:
$$\eqalignno{
\tilde I^{\rm(b1)}(\zeta)
&=\int_0^\infty \d s s^{-\zeta-1}
F_{00}^{\rm(b1)}(s,t) \cr
&={\Gamma(1-\zeta)\Gamma(1+\zeta)\over\Gamma(2)}\,
2\int_0^1 \rho \d\rho \int_0^1 \d x
\int_0^1 \d\alpha_1 \d\alpha_2 \d\alpha_3 \d\alpha_4\,
\delta(1-\sum\alpha_i) \cr
&\null\times n_1{N_{II}^{\hbox{\eightrm box}}(\beta)/s\over
\Lambda(\beta)^2} \,
(\beta_1\beta_2)^{\zeta-1}\, d^{-1-\zeta}, &(\nr)}$$
where $d$ is given by eq.~{\eqfourseven},
$$\beta_1\beta_2=\alpha_1\alpha_2\rho(1-\rho)x, \eqno(\nr)$$
and $n_1$ is given by {\eqmdef}.
We conclude that the dominant contribution comes when
$\alpha_1\simeq\alpha_2\simeq0$ and $x\simeq0$.
We are interested in this contribution, which is of order $\zeta^{-3}$,
and the subdominant one, which is of order $\zeta^{-2}$.
In the latter contribution, two of the conditions
$\alpha_1\simeq0$, $\alpha_2\simeq0$ and $x\simeq0$
must be fulfilled.
This implies that
$$\beta_1=\alpha_1r_1\simeq0, \qquad
\qquad\alpha_2r_2\simeq0, \qquad\hbox{and}
\qquad\alpha_1r_2\simeq0. \eqno(\nr)$$
Thus, performing a Cheng-Wu transformation, we get
$$\eqalignno{
\tilde I^{\rm(b1)}(\zeta)&=4e^{-i\pi\zeta}
\int_0^1\d\rho \int_0^1\d x \int_0^\infty\d\alpha_1
\int_0^\infty\d\alpha_2 \int_0^1\d \alpha_3 \int_0^1\d \alpha_4\,
\delta(1-\alpha_3-\alpha_4) \cr
&\null\times\rho^\zeta(1-\rho)^{1+\zeta}\,
{1\over 1+\alpha_2}
[\alpha_1\alpha_2x]^{\zeta-1}\, (-d)^{-1-\zeta}, &(\nr)}$$
where [cf.\ eq.~{\eqfourseven}]
$$\eqalignno{
-d&=\alpha_3\alpha_4
+[\alpha_2^2+(\alpha_2+\alpha_3+\alpha_4)\alpha_1\rho^2]m^2 \cr
&+[(\alpha_3+\alpha_4)+\alpha_1(1-\rho)](\alpha_2+\alpha_3+\alpha_4)\lambda^2.
&(\nr)}$$
The integration over $x$ gives a factor $1/\zeta$.
The integration over $\alpha_1$ is given by eq.~{\eqIntbeta},
$$
\int_0^\infty\d\alpha_1\, \alpha_1^{\zeta-1}[a\alpha_1+b]^{-1-\zeta}
={1\over\zeta}\, a^{-\zeta}\, b^{-1}. \eqno(\nr)$$

Introducing a new variable $z$ as
$$\alpha_3=z, \qquad \alpha_4=1-z, \eqno(\nr)$$
we obtain
$$\eqalignno{
\tilde I^{\rm(b1)}(\zeta)&={4\over\zeta^2}\, e^{-i\pi\zeta}
\int_0^1\d\rho\, \rho^\zeta(1-\rho)^{1+\zeta}
[\rho^2m^2+(1-\rho)\lambda^2]^{-\zeta} \cr
&\times\int_0^1\d z \int_0^\infty\d\alpha_2 \,
\alpha_2^{\zeta-1}(1+\alpha_2)^{-\zeta-1} \cr
&\times[z(1-z)+\alpha_2^2m^2+(1+\alpha_2)\lambda^2]^{-1}.
&(\nr)}$$
The integration over $\alpha_2$ is done by applying the theorem
{\eqTheoremtwo}
which correctly reproduces the leading and subleading terms.
We get
$$\tilde I^{\rm(b1)}(\zeta)={4\over\zeta^2}\, e^{-i\pi\zeta}
\biggl[{1\over\zeta}\, F_\rho\, C_{20}(\lambda^2)
+\half R(m^2,\lambda^2)\biggr],
\eqnref{\eqFR}$$
where the function $C_{20}$ is defined by eq.~{\eqCtwozero},
$$F_\rho=\int_0^1\d\rho\, \rho^\zeta(1-\rho)^{1+\zeta}
[\rho^2m^2+(1-\rho)\lambda^2]^{-\zeta}, \eqno(\nr)$$
and the remainder $R$ is the function defined by eq.~{\eqRdef}
and evaluated in appendix~D.
The factor $1/2$ in front of $R$ in eq.~{\eqFR}
comes from the $\rho$-integral.

We need $F_\rho$ to order $\zeta$.  Since, by definition
$$\int_0^1\d x\, x^{\mu-1}(1-x)^{\nu-1}=B(\mu,\nu)=B(\nu,\mu),
\eqnref{\eqBetatwo}$$
and furthermore
$$\int_0^1\d\rho(1-\rho)\log[\rho^2m^2+(1-\rho)\lambda^2]
=\half[\log(m^2)-3] +\Order(\lambda^2/m^2), \eqno(\nr)$$
we have
$$F_\rho=\half\{1-\zeta[\log(m^2)-1]\}. \eqno(\nr)$$
Collecting all terms, we get
$$\tilde I^{\rm(b1)}(\zeta)
={2\over\zeta^3}\, e^{-i\pi\zeta}
\bigl\{C_{20}(\lambda^2)
+\zeta\bigl[
-C_{20}(\lambda^2)\log(m^2) +C_{20}(\lambda^2)
+R(m^2,\lambda^2) \bigr]\bigr\}. \eqno(\nr)$$
\newsubsec{THE TERMS $F_{00}^{\rm(b2)}(s,t)$ and $F_{00}^{\rm(b3)}(s,t)$}
The contributions $F_{00}^{\rm(b2)}(s,t)$ and $F_{00}^{\rm(b3)}(s,t)$
are evaluated in appendix~E.
They can be discarded.
\newsubsec{THE TERM $F_{00}^{\rm(b4)}(s,t)$}
The amplitude corresponding to the $F_{00}^{\rm(b4)}$ of eqs.~{\eqGamma}
and {\eqFonethree} is given by
$${\cal M}_{\rm(b4)}=(-ie)^4(M^\gamma)_{\alpha\nu}(M_q)^{\alpha\nu},
\eqno(\nr)$$
with
$$\eqalignno{
(M^\gamma)_{\alpha\nu}
&=[\bar u(p_1')\gamma^\beta u(p_1)]\;
[\bar v(p_2) \gamma_\alpha\gamma_\nu\gamma_\beta v(p_2')], &(\nr) \cr
(M_q)^{\alpha\nu}
&={\alpha\over2\pi}\int_0^1\rho \d\rho \int_0^1 \d x \, f_4(\rho,x) \cr
&\null\times \int{\d^4q\over(2\pi)^4}\,
{p_1^\alpha(-p_2-q)^\nu \over D((p_1-q)^2, q^2) [(p_2+q)^2-m^2+\ieps]} \cr
&\null\times {1\over q^2-\lambda^2+\ieps} \,
{1\over (p_1-p_1'-q)^2-\lambda^2+\ieps}. &(\nr)}$$
Here [see eq.~{\eqGamma}]
$$f_4(\rho,x)=-2(1-\rho)(1-\rho x), \eqno(\nr)$$
and
$$D((p_1-q)^2, q^2)=
r_1[(p_1-q)^2-m^2] +r_2[q^2-\lambda^2]-M^2+\ieps, \eqno(\nr) $$
with $r_1$ and $r_2$ defined by eq.~{\eqmdef} and $M^2$ by eq.~{\eqMtwo}.
The integration over momenta is performed as in sect.~3.1 and gives
$$\eqalignno{
(M_q)^{\alpha\nu}
&={\alpha\over2\pi}\int_0^1\rho \d\rho \int_0^1 \d x \, f_4(\rho,x) \cr
&\null\times {i\over16\pi^2}
\int_0^1 \d\alpha_1\cdots \d\alpha_4\, \delta(1-\sum\alpha_i)\,
{1\over\Lambda(\gamma)} \,
{p_1^\alpha k_2^\nu\over D(\gamma)^2}, &(\nr)}$$
with
$$\eqqqalignno{
\gamma_1&=\alpha_1r_1,           & \gamma_2&=\alpha_2, \cr
\gamma_3&=\alpha_3+\alpha_1r_2,  & \gamma_4&=\alpha_4. &(\nr) }$$
Furthermore [cf.\ eqs.\ {\eqLambdab} and {\eqfourseven}],
$$\eqalignno{
\Lambda(\gamma)&=\gamma_1+\gamma_2+\gamma_3+\gamma_4, &(\nr) \cr
D(\gamma)&=D_{\hbox{\eightrm box}}(\gamma)
-\alpha_1\Lambda(\gamma)M^2 \cr
&\equiv \gamma_1\gamma_2 s +\tilde d, \eqalref{\eqDbfour} }$$
with $M^2$ given by eq.~{\eqMtwo},
$$M^2=\rho^2m^2 +(1-\rho)\lambda^2-r_2\lambda^2, \eqno(\nr)$$
and
$$k_2=-\gamma_1p_1-(\gamma_1+\gamma_3+\gamma_4)p_2-\gamma_4Q, \eqno(\nr)$$
where $Q=p_1-p_1'$.

The reduction of the spinor part of ${\cal M}_{\rm(b4)}$ is now
straightforward.
For the function $F_{00}^{\rm(b4)}(s,t)$, defined by the decomposition
{\eqFtwoloop}--{\eqDalpha}, we get
$$\eqalignno{
F_{00}^{\rm(b4)}(s,t)&=
\int_0^1\rho \d\rho\int_0^1 \d x \, f_4(\rho,x)
\int \d\alpha_1\cdots \d\alpha_4 \, \delta(1-\sum\alpha_i) \cr
&\null\times{1\over\Lambda(\gamma)} \,
{N_{II}\over D(\gamma)^2}, \eqalref{\eqFbfour} \cr
N_{II}&=-2s(\gamma_1+\gamma_3+\gamma_4).
&(\nr)}$$

A Mellin transform of eq.~{\eqFbfour} with $|t|=1$ gives
$$\eqalignno{
\tilde I^{\rm(b4)}(\zeta)
&=\int_0^\infty \d s\, s^{-\zeta-1}\, F_{00}^{\rm(b4)}(s,t) \cr
&=e^{-i\pi\zeta} {\Gamma(1-\zeta)\Gamma(1+\zeta)\over\Gamma(2)}\,
\int_0^1 \rho\d\rho \int_0^1\d x\ f_4(\rho,x) \cr
&\null\times \int_0^1 \d\alpha_1 \d\alpha_2 \d\alpha_3 \d\alpha_4\,
\delta(1-\sum\alpha_i) \cr
&\null\times {2(\gamma_1+\gamma_3+\gamma_4)\over \Lambda(\gamma)} \,
(r_1\alpha_1\alpha_2)^{-1+\zeta}\, (-\tilde d)^{-1-\zeta}, &(\nr)}$$
with $\tilde d$ given by eq.~{\eqDbfour} as
$$\eqalignno{
-\tilde d
&=\gamma_3\gamma_4+[(\gamma_1+\gamma_2)^2+\alpha_1\rho^2\Lambda(\gamma)]m^2
\cr
&+[\gamma_3+\gamma_4+\alpha_1(1-\rho)-\alpha_1 r_2]\Lambda(\gamma)\lambda^2.
\eqalref{\eqdbfour}}$$
We here use the Cheng-Wu theorem to rescale the $\alpha_1$ and
$\alpha_2$-variables,
$$\eqalignno{
\tilde I^{\rm(b4)}(\zeta)
&=2e^{-i\pi\zeta} \int_0^1 \rho\d\rho \int_0^1\d x\ f_4(\rho,x) \cr
&\null\times \int_0^\infty \d\alpha_1 \int_0^\infty \d\alpha_2
\int_0^1 \d\alpha_3 \int_0^1 \d\alpha_4\, \delta(1-\alpha_3-\alpha_4)
{\gamma_1+\gamma_3+\gamma_4 \over \Lambda(\gamma)} \cr
&\null\times [\rho(1-\rho)x\alpha_1\alpha_2]^{-1+\zeta}\,
(-\tilde d)^{-1-\zeta},
&(\nr)}$$
where we have also substituted for $r_1$ according to eq.~{\eqmdef},
and replaced $\Gamma(1-\zeta)\Gamma(1+\zeta)$ by unity.
The leading singularity is obtained when $\alpha_1\simeq\alpha_2\simeq0$
and $x\simeq0$.
We employ the singularity theorem of eq.~{\eqTheoremtwo},
$$\tilde I^{\rm(b4)}(\zeta)
=\tilde I_0(\zeta) +\tilde I_1(\zeta) +\tilde I_2(\zeta),
\eqnref{\eqIbfour}$$
where the first term corresponds to the leading singularity at
$\alpha_1\simeq\alpha_2\simeq0$, and $\tilde I_1(\zeta)$ and
$\tilde I_2(\zeta)$ to the residual contributions from integrations
over $\alpha_1$ and $\alpha_2$, respectively.

The dominant term is evaluated by setting $\alpha_1=\alpha_2=0$ everywhere
except in the factor $(\alpha_1\alpha_2)^{-1+\zeta}$,
$$\eqalignno{
\tilde I_0(\zeta)
&=-{4\over\zeta^2}e^{-i\pi\zeta}\int_0^1 \d\rho \int_0^1\d x(1-\rho x)
\int_0^1 \d\alpha_3 \int_0^1 \d\alpha_4\, \delta(1-\alpha_3-\alpha_4) \cr
&\null\times [\rho(1-\rho)]^\zeta\, x^{-1+\zeta}\,
[\alpha_3\alpha_4+\lambda^2]^{-1-\zeta}, &(\nr)}$$
where we have substituted for $f_4(\rho,x)$, and evaluated
$(\gamma_1+\gamma_3+\gamma_4)/\Lambda(\gamma)$ and $\tilde d$ for
$\alpha_1=\alpha_2=0$.
The integration over $x$ is trivial, and the one over $\alpha_3$ and $\alpha_4$
is done by means of eqs.~{\eqCtwozero} and {\eqCtwoone},
$$\eqalignno{
\tilde I_0(\zeta)
&=-{4\over\zeta^3}e^{-i\pi\zeta}\int_0^1 \d\rho [\rho(1-\rho)]^\zeta\,
(1-\zeta\rho) \cr
&\null\times [C_{20}(\lambda^2) -\zeta C_{21}(\lambda^2)], &(\nr)}$$
where terms of $\Order(1/\zeta)$ have been discarded.
The integral over $\rho$ is again trivial, we get
$$\eqalignno{
\tilde I_0(\zeta)&=-{4\over\zeta^3}e^{-i\pi\zeta}\, C_{20}(\lambda^2) \cr
&+{4\over\zeta^2}\biggl[{5\over2}C_{20}(\lambda^2) +C_{21}(\lambda^2) \biggr].
&(\nr)}$$

The contribution arising from the integration over $\alpha_1$ is
$$\eqalignno{
\tilde I_1(\zeta)
&={2\over\zeta} \int_0^1 \rho\d\rho \int_0^1\d x\ f_4(\rho,x)
[\rho(1-\rho)x]^{-1+\zeta} \cr
&\null\times \int_0^1 \d\alpha_3 \int_0^1 \d\alpha_4\,
\delta(1-\alpha_3-\alpha_4) \cr
&\null\times \biggl\{ \int_0^1{\d\alpha_1\over\alpha_1}
\biggl[{\gamma_1+\gamma_3+\gamma_4\over \Lambda(\gamma)}\bigg|_{\alpha_2=0}\,
{1\over a(\alpha_1,0)} -{1\over a(0,0)} \biggr] \cr
&+\int_1^\infty {\d\alpha_1\over\alpha_1}
{\gamma_1+\gamma_3+\gamma_4\over \Lambda(\gamma)}\bigg|_{\alpha_2=0}\,
{1\over a(\alpha_1,0)} \biggr\},
&(\nr)}$$
where we have used the notation [cf.~eq.~{\eqdbfour}]
$$a(\alpha_1,\alpha_2)=-\tilde d. \eqno(\nr)$$
One more power of $1/\zeta$ is required.  Since this can only come from the
integration over $x$, we may put $x=0$ everywhere except in the factor
$x^{-1+\zeta}$.  In this limit, the factor
$(\gamma_1+\gamma_3+\gamma_4)/\Lambda(\gamma)=1$.
With $x=0$, we have
$$\eqalignno{
a(\alpha_1,0)&=\alpha_3\alpha_4 +\alpha_1\rho^2m^2
+[1+\alpha_1(1-\rho)]\lambda^2 \cr
&\equiv A\alpha_1 +B, &(\nr)}$$
and obtain
$$\eqalignno{
\tilde I_1(\zeta)
&=-{4\over\zeta^2} \int_0^1 \d\rho
\int_0^1 \d\alpha_3 \int_0^1 \d\alpha_4\,
\delta(1-\alpha_3-\alpha_4) {1\over \alpha_3\alpha_4 +\lambda^2} \cr
&\null\times \biggl\{
-\int_0^1{\d\alpha_1\over\alpha_1+B/A}
+\int_1^\infty \d\alpha_1\biggl({1\over \alpha_1}
-{1\over\alpha_1+B/A}\biggr)\biggr\} \cr
&=-{4\over\zeta^2} \int_0^1 \d\rho
\int_0^1 \d\alpha_3 \int_0^1 \d\alpha_4\,
\delta(1-\alpha_3-\alpha_4) {1\over \alpha_3\alpha_4 +\lambda^2} \cr
&\null\times \biggl\{
\log[\alpha_3\alpha_4 +\lambda^2]
-\log[\rho^2m^2 +(1-\rho) \lambda^2] \biggr\}. &(\nr)}$$
Invoking the definitions {\eqCtwozero} and {\eqCtwoone}
we finally obtain
$$\eqalignno{
\tilde I_1(\zeta)
&=-{4\over\zeta^2} \int_0^1 \d\rho
\bigl\{C_{21}(\lambda^2)
-C_{20}(\lambda^2)\log[\rho^2m^2 +(1-\rho) \lambda^2]\bigr\} \cr
&=-{4\over\zeta^2} \bigl\{C_{21}(\lambda^2)
-C_{20}(\lambda^2)[\log m^2 -2]\bigr\}. &(\nr)}$$

Similarly, the contribution arising from the integration over $\alpha_2$ can be
expressed in terms of the function $R$ of appendix~D,
$$\tilde I_2(\zeta)=-{4\over\zeta^2} R(m^2,\lambda^2). \eqno(\nr)$$

Adding the three terms of eq.~{\eqIbfour} together,
the $C_{21}$-terms cancel and we get
$$\eqalignno{
\tilde I^{\rm(b4)}(\zeta)
&=-{4\over\zeta^3}\, e^{-i\pi\zeta}\, C_{20}(\lambda^2) \cr
&+{4\over\zeta^2}\biggl[(\half+\log m^2)C_{20}(\lambda^2)
-R(m^2,\lambda^2)\biggr].
&(\nr)}$$

\newsubsec{THE SUM OF TERMS}
Since the amplitudes
$F_{00}^{\rm(b2)}(s,t)$ and $F_{00}^{\rm(b3)}(s,t)$ are negligible,
we obtain for the Mellin transform of $F_{00}^{\rm (b)}(s,t)$
of eq.~{\eqFonethree}
the sum of the contributions from $\tilde I^{\rm(b1)}(\zeta)$ and
$\tilde I^{\rm(b4)}(\zeta)$,
and for $|t|=1$,
$$\eqalignno{
\tilde I^{\rm(b)}(\zeta)
&=-{2\over\zeta^3}\, e^{-i\pi\zeta}\, C_{20}(\lambda^2) \cr
&+{2\over\zeta^2}\biggl[(\log m^2+2)C_{20}(\lambda^2)
-R(m^2,\lambda^2)\biggr].
&(\nr)}$$

\newsec{Diagram (d)}
It is convenient to express the amplitudes for diagrams (d)
and ($\hbox{d}'$) in terms of
the renormalized electron self energy insertion.
\newsubsec{THE ONE-LOOP ELECTRON SELF ENERGY}
The expression for the renormalized one-loop
electron self energy $\Sigma(p)$ of fig.~5 is usually written as
$$\Sigma(p)=(\pslash-m)^2\bigl[(\pslash+m) \Sigma_1(p^2)
+m \Sigma_2(p^2)\bigr], \eqnref{\eqSigma}$$
with
$$\eqalignno{
\Sigma_1(p^2)
&={\alpha\over2\pi}\int_0^1 \d\rho \int_0^1 \d x
{\rho(1-\rho)^2 \over
x\rho(1-\rho)(p^2-m^2) -\rho^2m^2 -(1-\rho)\lambda^2 +\ieps} \cr
&\quad\times\biggl[1-{2x\rho(1+\rho)m^2\over \rho^2m^2+(1-\rho)\lambda^2}
\biggr], &(\nr) \cr
\Sigma_2(p^2)
&=-{\alpha\over2\pi}\int_0^1 \d\rho \int_0^1 \d x
{\rho(1-\rho^2) \over
x\rho(1-\rho)(p^2-m^2) -\rho^2m^2 -(1-\rho)\lambda^2 +\ieps} . &(\nr)}$$

We can rewrite $\Sigma(p)$ on a form more convenient for us,
$$\eqalignno{
\Sigma(p)
&=(\pslash-m)^2(\pslash+m)\Sigma_1^{(0)}(p^2)
+m (p^2-m^2)\Sigma_2(p^2) +(\pslash-m)A, \eqalref{\eqSigmadecomp}}$$
where
$$\Sigma_1^{(0)}(p^2)
={\alpha\over2\pi}\int_0^1 \d\rho \int_0^1 \d x
{\rho(1-\rho)^2 \over
x\rho(1-\rho)(p^2-m^2) -\rho^2m^2 -(1-\rho)\lambda^2 +\ieps}, \eqno(\nr)$$
and
$$A=-{\alpha\over\pi}\int_0^1\d \rho
{\rho(1-\rho^2)m^2\over \rho^2m^2+\lambda^2}
=-{\alpha\over2\pi}\biggl[\log\biggl({m^2\over\lambda^2}\biggr)-1\biggr].
\eqnref{\eqA}$$
Since $A$ is a constant the third term in eq.~{\eqSigmadecomp}
gives a contribution to $F_{00}^{\rm(d)}$ proportional to the box
diagram.
\newsubsec{DIAGRAM (d) WITH SELF-ENERGY INSERTION}
Diagram (d) of fig.~1 represents the self-energy correction
to the internal electron line.  The renormalized self-energy function
$\Sigma(p)$ is given above, eq.~{\eqSigmadecomp}.

We can write the amplitude $\calM{(d)}$ in terms of the one-loop
self-energy insertion $\Sigma(p)$ as
$$\eqalignno{
\calM{(d)}=&\int{\d^4q\over(2\pi)^4}
\bar u(p_1')(-ie\gamma^\beta)
{i(\rlap/ p_1-\rlap/ q +m) \over (p_1-q)^2-m^2+i\epsilon} \cr
\times&[-i\Sigma(p_1-q)]
{i(\rlap/ p_1-\rlap/ q +m) \over (p_1-q)^2-m^2+i\epsilon}
(-ie\gamma^\alpha) u(p_1) \cr
\times&\bar v(p_2)(-ie\gamma_\alpha)
{i(-\rlap/ p_2-\rlap/ q +m) \over (p_2+q)^2-m^2+i\epsilon}
(-ie\gamma_\beta) v(p_2') \cr
\times&{-i\over q^2-\lambda^2+i\epsilon}\;
{-i\over (p_1-p_1'-q)^2-\lambda^2+i\epsilon}. &(\nr)}$$
Here, we expand $F_{00}^{\rm(d)}(s,t)$ as a sum of three terms,
$$F_{00}^{\rm(d)}(s,t)=F_{00}^{\rm(d1)}(s,t)+F_{00}^{\rm(d2)}(s,t)
+F_{00}^{\rm(d3)}(s,t), \eqnref{\eqFdsum}$$
according to the decomposition {\eqSigmadecomp}.
The second term which comes from the $\Sigma_2$-part
is explicitly proportional to $m$ and
hence can be neglected.
Furthermore, the third part is trivial,
$${\alpha\over4\pi}\, F_{00}^{\rm(d3)}(s,t)
=AG_{00}^{\hbox{\eightrm box}}(s,t), \eqno(\nr)$$
where $A$ is given by eq.~{\eqA} and $G_{00}^{\hbox{\eightrm box}}(s,t)$
by eq.~{\eqGboxBhabha}.  We obtain
$$F_{00}^{\rm(d3)}(s,t)={8\over t}\, \log\biggl({s\over|t|}\biggr)\,
\log\biggl({|t|\over\lambda^2}\biggr)
\biggl[\log\biggl({m^2\over\lambda^2}\biggr)-1\biggr]. \eqnref{\eqFdthree}$$
This corresponds, for $|t|=1$, to the Mellin transform
$$\tilde I^{\rm(d3)}(\zeta)
=-{4\over\zeta^2}\, C_{20}(\lambda^2)
\biggl[\log\biggl({m^2\over\lambda^2}\biggr)-1\biggr].
\eqnref{\eqIdthree}$$

For the first part, originating from $\Sigma_1^{(0)}$, we have
$$\eqalignno{
F_{00}^{\rm(d1)}(s,t)&=
\int_0^1 \d\rho \int_0^1 \d x\,
2\,\rho(1-\rho)^2 \cr
&\null\times \int_0^1\ldots\int_0^1 \d\alpha_1\ldots \d\alpha_4
\delta(1-\sum\alpha_i){1\over \Lambda(\beta)^2} \cr
&\null\times\biggl[
{N_{II}^{\hbox{\eightrm box}}(\beta)\over D(\beta)^2}
+{N_{I}^{\hbox{\eightrm box}}(\beta)\over 2D(\beta)} \biggr]. &(\nr)}$$
Here $N_{II}^{\hbox{\eightrm box}}(\beta)$,
$N_I^{\hbox{\eightrm box}}(\beta)$ and $\Lambda(\beta)$
refer to the box-diagram
results of sect.~3.  Furthermore, in analogy with the analysis
of sect.~5, the transformed Feynman parameters are given by
$$\eqqqalignno{
\beta_1&=\alpha_1x\rho(1-\rho), & \beta_2&=\alpha_2, \cr
\beta_3&=\alpha_3,              & \beta_4&=\alpha_4, &(\nr)}$$
and the denominator function by
$$\eqalignno{
D(\beta)&=D_{\hbox{\eightrm box}}(\beta)
-\Lambda(\beta)\alpha_1[\rho^2 m^2+(1-\rho)\lambda^2] \cr
&=\beta_1\beta_2s+\beta_3\beta_4t
-[(\beta_1+\beta_2)^2+\Lambda(\beta)\alpha_1\rho^2]m^2 \cr
&\quad-[(\beta_3+\beta_4)+\alpha_1(1-\rho)]\Lambda(\beta)\lambda^2
+\ieps \cr
&\equiv\beta_1\beta_2s +d, \eqalref{\eqDd}}$$
where
$$\eqalignno{
\Lambda(\beta)&=\beta_1+\beta_2+\beta_3+\beta_4 \cr
&=\alpha_1x\rho(1-\rho)+\alpha_2+\alpha_3+\alpha_4. &(\nr)}$$

The contribution from the $N_{I}^{\hbox{\eightrm box}}$-term
is of order $1/s$, with the leading term being proportional to $\log^3(s)/s$,
and may be neglected.  Thus,
$$\eqalignno{
F_{00}^{\rm(d1)}(s,t)
&=2\int_0^1 \d\rho \int_0^1 \d x\,
\rho(1-\rho)^2 \cr
&\null\times \int_0^1\ldots\int_0^1 \d\alpha_1\ldots \d\alpha_4 \,
\delta(1-\sum\alpha_i){1\over \Lambda(\beta)^2} \,
{N_{II}^{\hbox{\eightrm box}}(\beta)\over D(\beta)^2}. \eqalref{\eqGone}}$$
where $N_{II}^{\hbox{\eightrm box}}(\beta)$ is given by eq.~{\eqNIIbox}.
\newsubsec{THE TERM $F_{00}^{\rm(d1)}$}
The Mellin transform of $F_{00}^{\rm(d1)}(s,t)$, with $|t|=1$, becomes
$$\eqalignno{
\tilde I^{\rm(d1)}(\zeta)&=\int_0^\infty \d s s^{-\zeta-1}\,
F_{00}^{\rm(d1)}(s,t) \cr
&=2{\Gamma(1-\zeta)\Gamma(1+\zeta)\over\Gamma(2)}
\int_0^1 \d\rho \rho(1-\rho)^2 \int_0^1 \d x
\int_0^1 \d\alpha_1\cdots \int_0^1 \d\alpha_4\, \delta(1-\sum\alpha_i)\cr
&\times{N_{II}^{\hbox{\eightrm box}}(\beta)/s\over\Lambda(\beta)^2}\,
[\alpha_1\alpha_2x\rho(1-\rho)]^{-1+\zeta}
\, d^{-1-\zeta}, &(\nr)}$$
with $d$ given by eq.~{\eqDd} and with $|t|=1$ as
$$\eqalignno{
d&=-\alpha_3\alpha_4
-\{[\alpha_1x\rho(1-\rho)+\alpha_2]^2
+\alpha_1\rho^2\Lambda(\beta)\}m^2 \cr
&\quad-[\alpha_3+\alpha_4+\alpha_1(1-\rho)]\Lambda(\beta)\lambda^2+\ieps,
&(\nr)}$$
where
$$\Lambda(\beta)=\alpha_1x\rho(1-\rho)+\alpha_2+\alpha_3+\alpha_4.
\eqno(\nr)$$

We first perform a Cheng-Wu transformation, letting $\alpha_1$ and $\alpha_2$
go to infinity.  Then, we rescale the $\alpha_1$ variable,
$$\alpha_1\to{\alpha_1\over x\rho(1-\rho)}, \eqno(\nr)$$
and exploit Theorem~1 of eq.~{\eqTheoremone}.
We get
$$\eqalignno{
\tilde I^{\rm(d1)}(\zeta)
&=-2e^{-i\pi\zeta}
\int_0^\infty\d\alpha_1 \int_0^\infty\d\alpha_2
\int_0^1 \d\alpha_3 \int_0^1 \d\alpha_4\, \delta(1-\alpha_3-\alpha_4) \cr
&\times\int_0^1 \d\rho (1-\rho) \int_0^1 {\d x\over x}
{N_{II}^{\hbox{\eightrm box}}(\alpha)/s\over\Lambda(\alpha)^2}\,
(\alpha_1\alpha_2)^{-1+\zeta}\biggl[a+{b\over x}\biggr]^{-1-\zeta},
&(\nr)}$$
with
$$\eqalignno{
a&=\alpha_3\alpha_4+(\alpha_1+\alpha_2)^2m^2
+\Lambda(\alpha)(\alpha_3+\alpha_4)\lambda^2 >0, \eqalref{\eqalower} \cr
b&={\alpha_1\Lambda(\alpha)\over\rho(1-\rho)}
[\rho^2m^2+(1-\rho)\lambda^2] >0, &(\nr)}$$
and
$$\Lambda(\alpha)=\alpha_1+\alpha_2+\alpha_3+\alpha_4. \eqno(\nr)$$
The $x$-integral becomes
$$\eqalignno{
G_x&=\int_0^1{\d x\over x}\, {1\over [a+(b/x)]^{1+\zeta}} \cr
&=a^{-1-\zeta}\,
\int_0^1 \d x\, {x^\zeta\over [x+(b/a)]^{1+\zeta}}. &(\nr)}$$
We expand the integrand in powers of $\zeta$ and get
$$G_x=a^{-1-\zeta}\,
\biggl\{ \log\biggl(1+{1\over\xi}\biggr)
+\sum_{n=1}^\infty {\zeta^n\over n!}
\int_0^{1/\xi} {\d t\over 1+t}\log^n\biggl({t\over 1+t}\biggr)\biggr\},
\eqno(\nr)$$
with $\xi=b/a$.  For the terms in the sum which contain extra powers of
$\zeta$, we can put $\alpha_1=\alpha_2=0$.
Since then $\xi=0$ we obtain
$$\int_0^\infty{\d t\over 1+t}\, \log^n\biggl({t\over 1+t}\biggr)
=n! (-1)^n\, S_{1,n}(1), \eqno(\nr)$$
where $S_{1,n}(1)$ is Nielsen's generalized polylogarithm
\ref{\refKolbig}{K.~S.\ K\"olbig, J.~A.\ Mignaco, and E.\ Remiddi,
BIT 10 (1970) 38},
\ref{\refDevoto}{A.\ Devoto and D.~W.\ Duke,
La Rivista del Nuovo Cimento 7 (1984) 1}.
Hence, the leading term of the sum, $n=1$, will contribute to
$\tilde I^{\rm(d1)}$ a term which is proportional to $\zeta^{-1}$, and
to $F_{00}^{\rm(d1)}$ a constant.
Such a term can be discarded, since it will cancel against a corresponding
contribution from the crossed diagram, $F_{00}^{\rm(d')}$.
We are left with
$$\eqalignno{
\tilde I^{\rm(d1)}(\zeta)
&=-2e^{-i\pi\zeta}
\int_0^\infty \d\alpha_1 \int_0^\infty \d\alpha_2
\int_0^1 \d\alpha_3\, \int_0^1 \d\alpha_4\, \delta(1-\alpha_3-\alpha_4)
\int_0^1 \d\rho(1-\rho) \cr
&\times {N_{II}^{\hbox{\eightrm box}}(\alpha)/s\over\Lambda(\alpha)^2}\,
(\alpha_1\alpha_2)^{-1+\zeta}
a^{-1-\zeta}\, \log\biggl({a+b\over b}\biggr). \eqalref{\eqstar}}$$
We split the logarithm of eq.~{\eqstar} into two parts,
$$\log\biggl({a+b\over b}\biggr)
=\log\biggl({a+b\over b/\alpha_1}\biggr) -\log\alpha_1, \eqno(\nr)$$
and denote the corresponding contributions to $\tilde I^{\rm(d1)}$ by
$\tilde I_a$ and $\tilde I_b$,
$$\tilde I^{\rm(d1)}(\zeta)=\tilde I_a(\zeta)+\tilde I_b(\zeta). \eqno(\nr)$$

For $\tilde I_a$ we find the dominant contribution
to come from $\alpha_1\simeq\alpha_2\simeq0$,
$$\eqalignno{
\tilde I_a(\zeta)
&=-{2\over \zeta^2}\, \int_0^1 \d\rho(1-\rho)
\int_0^1 \d\alpha_3\int_0^1 \d\alpha_4\,\delta(1-\alpha_3-\alpha_4) \cr
&\times \biggl[{N_{II}^{\hbox{\eightrm box}}(\alpha)/s\over\Lambda(\alpha)^2}\,
{1\over a}\,
\log\biggl({a+b\over b/\alpha_1}\biggr)\biggr]_{\alpha_1=\alpha_2=0} \cr
&={4\over \zeta^2}\, \int_0^1 \d\rho(1-\rho)
\int_0^1 \d\alpha_3\int_0^1 \d\alpha_4\,\delta(1-\alpha_3-\alpha_4) \cr
&\times {1\over \alpha_3\alpha_4+\lambda^2}\,
\log{[\alpha_3\alpha_4+\lambda^2]\rho(1-\rho) \over
\rho^2m^2+(1-\rho)\lambda^2} \cr
&={2\over \zeta^2}\,
\Bigl[C_{21}(\lambda^2) +SC_{20}(\lambda^2)\Bigr], \eqalref{\eqIonetilde}}$$
with
$$\eqalignno{
S&=-2\int_0^1 \d\rho(1-\rho)
\log\biggl({\rho^2m^2+(1-\rho)\lambda^2\over\rho(1-\rho)}\biggr) \cr
&=1-\log m^2. &(\nr)}$$

The next integral is $\tilde I_b$.
After integration over $\rho$, and symmetrizing in $\alpha_1$ and $\alpha_2$,
it is given by
$$\eqalignno{
\tilde I_b(\zeta)&=\half\,e^{-i\pi\zeta}
\int_0^\infty\d\alpha_1 \int_0^\infty\d\alpha_2
\int_0^1\d\alpha_3 \int_0^1\d\alpha_4 \,
\delta(1-\alpha_3-\alpha_4) \cr
&\null\times{N_{II}^{\hbox{\eightrm box}}(\alpha)/s\over\Lambda(\alpha)^2}\,
(\alpha_1\alpha_2)^{-1+\zeta}\,
{1\over a^{1+\zeta}}\, \log(\alpha_1\alpha_2). &(\nr)}$$
We have, with $\alpha_3+\alpha_4=1$,
$${N_{II}^{\hbox{\eightrm box}}(\alpha)\over
s}=-[2+2(\alpha_1+\alpha_2)+3\alpha_1\alpha_2]. \eqno(\nr)$$
Here, the term proportional to $\alpha_1\alpha_2$ does not contribute
to either of the two leading orders when
$\zeta\to0$, and will be ignored.
We introduce new variables,
$$\rho=\alpha_1+\alpha_2, \qquad \alpha_1=\rho x,
\qquad \alpha_2=\rho(1-x). \eqno(\nr)$$
The integration over the new $x$ variable is easily done,
with the aid of the formulas
$$\eqalignno{
\int_0^1 \d x[x(1-x)]^{-1+\zeta}
&=B(\zeta,\zeta)
={2\over\zeta}+\Order(\zeta), &(\nr) \cr
\int_0^1 \d x[x(1-x)]^{-1+\zeta}\,\log x
&=B(\zeta,\zeta)[\psi(\zeta)-\psi(2\zeta)] \cr
&=-{1\over\zeta^2} -{\pi^2\over6}+\Order(\zeta). &(\nr)}$$
Since $\Lambda(\alpha)=1+\rho$ we get
$$\eqalignno{
\tilde I_b(\zeta)&=-e^{-i\pi\zeta}
\int_0^1\int_0^1 \d\alpha_3 \d\alpha_4\,\delta(1-\alpha_3-\alpha_4)
\int_0^\infty \d\rho\,\rho^{-1+2\zeta} \cr
&\times{1\over(1+\rho)}\,{1\over a^{1+\zeta}} \,
\biggl[-{2\over\zeta^2}+ {4\over\zeta}\log\rho -{\pi^2\over3}\biggr],
\eqalref{\eqIoneb}}$$
where $a=a(\rho,\alpha_3,\alpha_4)$ is obtained from eq.~{\eqalower} as
$$a(\rho,\alpha_3,\alpha_4)=\alpha_3\alpha_4+\rho^2m^2
+(1+\rho)(\alpha_3+\alpha_4)\lambda^2 >0. \eqnref{\eqarho}$$
We neglect terms that contribute only to the final constant.
That means that the term $\pi^2/3$ can be neglected.

We now consider the $\rho$-integral.  We have
$$\eqalignno{
\int_0^1 \d\rho\,\rho^{-1+2\zeta}\,G(\rho)\, \log\rho
&=\int_0^1 \d\rho\,\rho^{-1+2\zeta}\,G(0) \,\log\rho
+\int_0^1 \d\rho\,\rho^{-1+2\zeta}[G(\rho)-G(0)] \,\log\rho \cr
&=-{1\over4\zeta^2}\,G(0), &(\nr)}$$
where the second integral in the intermediate step is finite when $\zeta\to0$
and has been discarded.  The first term of eq.~{\eqIoneb} is evaluated
with the aid of theorem~2. We get
$$\tilde I_b(\zeta)=-e^{-i\pi\zeta}
\biggl[-{2\over\zeta^3}X -{2\over\zeta^2}R(m^2,\lambda^2)\biggr], \eqno(\nr)$$
with
$$X=\int_0^1\int_0^1 \d\alpha_3 \d\alpha_4\,\delta(1-\alpha_3-\alpha_4)
{1\over a_0^{1+\zeta}}, \eqno(\nr) $$
where $a_0=a(\rho=0,\alpha_3,\alpha_4)$ of eq.~{\eqarho}
and where $R$ is the function evaluated in appendix D.
{}From the defining equations {\eqCtwozero} and {\eqCtwoone},
we get
$$X=C_{20}(\lambda^2)-\zeta C_{21}(\lambda^2)+\Order(\zeta^2). \eqno(\nr)$$
Collecting terms, we obtain
$$\tilde I_b(\zeta)
={2\over\zeta^3}\,e^{-i\pi\zeta}\,
\biggl\{C_{20}(\lambda^2)+\zeta
\biggl[-C_{21}(\lambda^2) +R(m^2,\lambda^2)\biggr]
\biggr\}. \eqnref{\eqIonebtilde} $$

Adding $\tilde I_a$ of eq.~{\eqIonetilde}
and $\tilde I_b$ of eq.~{\eqIonebtilde} together we get
$$\tilde I^{\rm(d1)}(\zeta)={2\over\zeta^3}\,e^{-i\pi\zeta}\,
\biggl\{C_{20}(\lambda^2)
+\zeta\biggl[(1-\log(m^2))C_{20}(\lambda^2)
+R(m^2,\lambda^2) \biggr] \biggr\}. \eqnref{\eqIdone}$$

\newsubsec{THE SUM OF TERMS FOR THE (d) DIAGRAM}
Since the term $F_{00}^{\rm(d2)}(s,t)$
can be neglected in our approximation we get,
adding the contributions {\eqIdthree} and {\eqIdone},
being the Mellin transforms of $F_{00}^{\rm(d3)}(s,t)$ and
$F_{00}^{\rm(d1)}(s,t)$, respectively, for $|t|=1$,
$$\eqalignno{
\tilde I^{\rm(d)}(\zeta)
&=\tilde I^{\rm(d1)}(\zeta) + \tilde I^{\rm(d3)}(\zeta) \cr
&={2\over\zeta^3}\, e^{-i\pi\zeta}\, C_{20}(\lambda^2) \cr
&+{2\over\zeta^2}\biggl\{\biggl[
-3(\log m^2-1) +2\log\lambda^2 \biggr] C_{20}(\lambda^2)
+R(m^2,\lambda^2) \biggr\}. &(\nr)}$$
\newsec{Summary}
The results for the Mellin transforms of the Feynman amplitudes
corresponding to the diagrams of fig.~1 and for $|t|=1$,
are as follows; for diagram a,
$$\tilde I^{\rm(a)}(\zeta)
={2\over\zeta^3}\, e^{-i\pi\zeta}\, C_{20}(\lambda^2)
+{2\over\zeta^2}\, R(m^2,\lambda^2), \eqno(\nr)$$
for the sum of diagrams b and c,
$$\eqalignno{
2\tilde I^{\rm(b)}(\zeta)
&=-{4\over\zeta^3}\, e^{-i\pi\zeta}\, C_{20}(\lambda^2) \cr
&+{4\over\zeta^2} \biggl\{
\bigl[\log m^2 +2\bigr] C_{20}(\lambda^2) -R(m^2,\lambda^2) \biggr\}, &(\nr)}$$
and finally, for diagram d,
$$\eqalignno{
\tilde I^{\rm(d)}(\zeta)
&={2\over\zeta^3}\, e^{-i\pi\zeta}\, C_{20}(\lambda^2) \cr
&+{2\over\zeta^2}\biggl\{\biggl[
-3(\log m^2-1) +2\log\lambda^2 \biggr] C_{20}(\lambda^2)
+R(m^2,\lambda^2) \biggr\}. &(\nr)}$$
The corresponding amplitudes, $F_{00}(s,t)$,
can be obtained from the above Mellin transforms
by reintroducing the scale factor $|t|$, and making the replacements
$$\eqalignno{
{1\over\zeta^3} &\to -{1\over2t}\,\log^2 {s\over|t|}, &(\nr) \cr
{1\over\zeta^2} &\to -{1\over t}\, \log {s\over|t|}, &(\nr) }$$
as described in appendix~A.
Thus, individual diagrams contain terms that are both linear and quadratic
in $\log(s/|t|)$.

Adding the Mellin transforms of the four diagrams (a), (b), (c) and (d),
we see that the terms proportional to $\zeta^{-3}$ cancel,
giving the sum
$$\tilde I(\zeta)
={2\over\zeta^2}\bigl[
-\log m^2 +7 +2\log\lambda^2\bigr] C_{20}(\lambda^2). \eqno(\nr)$$
Inverting the Mellin transform, and reinserting for $t$, we have
for the sum of amplitudes
$$F_{00}(s,t)={4\over t}\log{s\over|t|}
\biggl[\log{\lambda^2\over m^2} +\log{\lambda^2\over|t|} +7 \biggr]
\log{\lambda^2\over|t|}. \eqno(\nr)$$

We now add in the contributions from the four crossed diagrams
$({\rm a}')$, $({\rm b}')$, $({\rm c}')$ and $({\rm d}')$ of fig.~1
according to eqs.~{\eqcrossed} and {\eqtvaafem},
$$F_{00}(s,t)+F_{00}^{\hbox{\eightrm (cr)}}(s,t)
=-{4i\pi\over t}
\biggl[\log{\lambda^2\over m^2} +\log{\lambda^2\over|t|} +7 \biggr]
\log{\lambda^2\over|t|}. \eqno(\nr)$$

Thus, when we add the amplitudes for the four diagrams (a)--(d)
the terms proportional to $\log^2s$ cancel, and when in addition
the amplitudes for the four crossed diagrams $({\rm a}')$--$({\rm d}')$
are added, also the terms proportional to $\log s$ cancel.
The cancellation of terms that depend on $\log s$ has been
noted before (see ref.~{\refCW}, chapter~11).

Let us denote by ${\cal M}^{(3\gamma)}$ the summed contributions
from the eight diagrams in fig.~1, and by ${\cal M}^{(2\gamma)}$ the sum of
the box diagram and the crossed box diagram.
Then our result can be written as
$${\cal M}^{(3\gamma)}=
{\alpha\over4\pi}\biggl[
\log{\lambda^2\over m^2} +\log{\lambda^2\over|t|} +7 \biggr]
{\cal M}^{(2\gamma)}. \eqno(\nr)$$
If we compare the factor multiplying ${\cal M}^{(2\gamma)}$ with the electric
form factor, $F_1(t)$ of eq.~(B.30), we conclude that
${\cal M}^{(3\gamma)}\ne F_1(t) {\cal M}^{(2\gamma)}$.
Thus, we do not reproduce the result obtained in the eikonal approximation
{\refCW},
\ref{\refeikonal}
{H. Cheng and T.~T. Wu, Phys.\ Rev.\ 184 (1969) 1868; \lbrk
S.~J. Chang, Phys.\ Rev.\ D1 (1970) 2977; \lbrk
Y.~P. Yao, Phys.\ Rev.\ D3 (1971) 1364}.
The reason for this discrepancy must be sought in the linearization
of the denominators of the fermion propagators that is part of the
eikonal approximation.

There are other sets of Feynman diagrams still to
be evaluated before the virtual corrections are complete.
Work on the box diagrams with a photon self-energy insertion
is in progress and will be reported elsewhere.

\acknowledgment{
This research has been supported by the Research Council of Norway
and by the Swedish National Research Council.}
\vfill\eject

\newappendixtext{A}{The Mellin transform}
We start from integrals of the type
$$I(s)=\int\d^7\alpha\, F(\alpha)\, {1\over (As+B+\ieps)^a}, \eqno(\nr)$$
where only the $s$-dependence has been made explicit.
We assume that for large values of $s$,
$$I(s)={1\over s^n}\sum_{k=0}^i\, a_k (\log s)^k +\Order(1/s^{n+1}).
\eqnref{\eqMellintwo} $$
To calculate $a_0$, $\ldots$, $a_i$ we introduce the Mellin transform
\ref{\refBj}
{J.~D.\ Bjorken and T.~T.\ Wu, Phys.\ Rev.\ 130 (1963) 2566; \lbrk
see also P. Osland and T.~T.\ Wu, Nucl.\ Phys.\ B288 (1987) 77},
$$\eqalignno{
\tilde I(\zeta)
&=\int_0^\infty \d s\, s^{-\zeta+(n-1)}\, I(s) \cr
&={\Gamma(n-\zeta)\Gamma(a+\zeta-n)\over \Gamma(a)}
\int\d^7\alpha\, F(\alpha)\, (A+\ieps)^{\zeta-n} (B+\ieps)^{-a-\zeta+n}.
&(\nr)}$$
Then, we determine the structure of $\tilde I(\zeta)$ in the vicinity
of $\zeta=0$. From it we obtain the coefficients of eq.~{\eqMellintwo} as
$$\tilde I(\zeta)=\sum_{k=0}^i {k!\over \zeta^{k+1}}\, a_k
+(\hbox{regular terms at \ }\zeta=0). \eqno(\nr)$$

\newappendixtext{B}{The one-loop vertex function}
\noindent B.1\ GENERAL FORM OF THE VERTEX FUNCTION

The renormalized vertex function of fig.~6 can be expanded as
$$\eqalignno{
\Gamma_\mu(p',p;q)&=A(\pslash'-m) \gamma_\mu(\pslash-m)
+(\pslash'-m)B_\mu' +B_\mu(\pslash-m) +C_\mu\,, \eqalref{\eqBone} \cr
q&=p-p'. &(\nr)}$$

In order to be able to give the expressions for the coefficients
in a compact form, we introduce the notation
$$F[g(\alpha)]={\alpha\over2\pi}
\int_0^1\ldots\int_0^1 \d\alpha_1 \d\alpha_2 \d\alpha_3
\delta(1-\sum\alpha_i)\, {g(\alpha)\over D(\alpha)}, \eqno(\nr)$$
with
$$\eqalignno{
D(\alpha)&=\alpha_1\alpha_3(p'{}^2-m^2)
+\alpha_2\alpha_3(p^2-m^2) +\alpha_1\alpha_2q^2 +D_0, &(\nr) \cr
D_0&=-(\alpha_1+\alpha_2)^2m^2 -\alpha_3\lambda^2 +\ieps, &(\nr)}$$
We observe that $F[g(\alpha)]$ is invariant under the simultaneous
interchanges $\alpha_1\leftrightarrow\alpha_2$,
and $p'{}^2\leftrightarrow p^2$.  Thus, for example
$$F[\alpha_3(1-\alpha_2)]=F[\alpha_3(1-\alpha_1)]
\Big|_{p'{}^2\leftrightarrow p^2}. \eqno(\nr)$$

The coefficients of eq.~{\eqBone} may now be expressed as
$$\eqalignno{
A&=F[\alpha_3], &(\nr) \cr
B_\mu'&=mF[\alpha_3]\gamma_\mu
+F[\alpha_3(1-\alpha_1)]\{(\pslash'+m)\gamma_\mu-2p_\mu\} \cr
&-F[2\alpha_1(1-\alpha_1)]q_\mu, &(\nr) \cr
B_\mu&=mF[\alpha_3]\gamma_\mu
+F[\alpha_3(1-\alpha_2)]\{\gamma_\mu(\pslash+m)-2p_\mu'\} \cr
&+F[2\alpha_2(1-\alpha_2)]q_\mu, &(\nr) \cr
C_\mu&=\gamma_\mu\{\GR +m^2F[-1+2\alpha_3+\alpha_3^2]
-q^2F[(1-\alpha_1)(1-\alpha_2)]-\tilde C\} \cr
&+2m\{p_\mu'F[1-2\alpha_1-\alpha_3(1-\alpha_1)]
     +p_\mu F[1-2\alpha_2-\alpha_3(1-\alpha_2)]\}, &(\nr) \cr
\tilde C&=m^2F[-1 +4\alpha_3 -\alpha_3^2]\Big|_{p'{}^2=p^2=m^2, \,\,q^2=0}
\eqalref{\eqCtilde} }$$
where the function $\GR$ is a renormalized one, and given by
$$\eqalignno{
\GR
&=-{\alpha\over2\pi}
\int_0^1\ldots\int_0^1 \d\alpha_1 \d\alpha_2 \d\alpha_3
\delta(1-\sum\alpha_i) \cr
&\null\times\bigl\{\log(D(\alpha)) -\log[-(\alpha_1+\alpha_2)^2m^2
-\alpha_3\lambda^2 +\ieps]\bigr\}. \eqalref{\eqGR} }$$
Similarly, $\tilde C$ of eq.~{\eqCtilde} is a subtraction constant
that ensures the proper renormalization condition,
$$\bar u(p')\Gamma_\mu(p'{}^2=m^2,\, p^2=m^2;\, q^2=0)u(p)=0.
\eqno(\nr)$$

Furthermore, we can introduce new variables
$\alpha_1=\rho x$ and $\alpha_2=\rho(1-x)$ and write
$$\eqalignno{
F[g(\alpha)] &={\alpha\over2\pi}\int_0^1\rho \d\rho \int_0^1 \d x
{g(\alpha)\over D(\rho,x)}, &(\nr) \cr
D(\rho,x)&=\rho(1-\rho)x(p'{}^2-m^2)
+\rho(1-\rho)(1-x)(p^2-m^2) +\rho^2x(1-x)q^2 \cr
&+D_0, \eqalref{\eqDrhox} \cr
D_0&=-\rho^2m^2 -(1-\rho)\lambda^2+\ieps, &(\nr) }$$
where it is understood that the function $g(\alpha)$
is to be expressed in terms of $\rho$ and $x$.

\vskip8pt\goodbreak
\noindent B.2\ HALF-OFF-SHELL VERTEX FUNCTION

When one particle is on mass shell, e.g.,
$p^2=m^2,$
we can write $\GR$ of eq.~{\eqGR} after partial integration, as
$$\eqalignno{
\GR(p^2=m^2,p'{}^2,q^2)
=-{\alpha\over2\pi}
\biggl\{&(p'{}^2-m^2)\int_0^1\rho \d\rho \int_0^1 \d x
{\rho(1-\rho)(1-x)\over D(p'{}^2,q^2)} \cr
-&q^2\int_0^1\rho \d\rho \int_0^1 \d x
{(1-x)(2x-1)\rho^2\over D(p'{}^2,q^2)} \biggr\}, &(\nr) }$$
with $D(p'{}^2,q^2)$ obtained from eq.~{\eqDrhox} by putting $p^2=m^2$,
$$D(p'{}^2,q^2)=\rho(1-\rho)x(p'{}^2-m^2)+\rho^2x(1-x)q^2+D_0+\ieps,
\eqno(\nr)$$
so that with $p^2=m^2$,
$$\eqalignno{
\Gamma_\mu(p',p;q)
&=(\pslash'-m)B_\mu'+C_\mu \cr
&=\gamma_\mu \bigl\{
\FR(p^2=m^2,p'{}^2,q^2) +m^2F[-1+2\alpha_3+\alpha_3^2]-\tilde C \bigr\} \cr
&+(\pslash'-m)\bigl\{-2p_\mu F[\alpha_3(1-\alpha_1)]
-q_\mu F[2\alpha_1(1-\alpha_1)]
+m\gamma_\mu F[\alpha_3]\bigr\}\cr
&+2m\bigl\{p_\mu'F[1-2\alpha_1-\alpha_3(1-\alpha_1)]
+p_\mu F[1-2\alpha_2-\alpha_3(1-\alpha_2)]\bigr\}. \eqalref{\eqhalfoff}}$$

Here,
$$\eqalignno{
\FR(p^2=m^2,p'{}^2,q^2)&=
\GR(p^2=m^2,p'{}^2,q^2)+(p'{}^2-m^2)F[\alpha_3(1-\alpha_1)] \cr
& -q^2F[(1-\alpha_1)(1-\alpha_2)] &(\nr) \cr
&={\alpha\over2\pi} \int_0^1 \rho \d\rho \int_0^1 \d x \,
{1\over D(p'{}^2,q^2)} \cr
&\null\times \bigl[n_1(p'{}^2-m^2) +n_2(q^2-\lambda^2) \bigr], &(\nr) }$$
with
$$\eqalignno{
D(p'{}^2,q^2)&\equiv r_1(p'{}^2-m^2) +r_2(q^2-\lambda^2) -M^2
+\ieps, &(\nr) \cr
M^2&=\rho^2m^2 +(1-\rho)\lambda^2-r_2\lambda^2, &(\nr)}$$
and
$$\eqqqalignno{
n_1&=(1-\rho)^2,   & n_2&=-(1-\rho)-\rho^2(1-x)^2, &(\nr) \cr
r_1&=\rho(1-\rho)x, & r_2&=\rho^2x(1-x). &(\nr) }$$

\vskip8pt\goodbreak
\noindent B.3\ ON-SHELL VERTEX FUNCTION

When both fermions are on-shell, $p^2=p'{}^2=m^2$, we have
$$\bar u(p')\Gamma_\mu(p',p;q) u(p)
=\bar u(p')\biggl[\gamma_\mu F_1(q^2)
-{i\over 2m}\sigma_{\mu\nu}q^\nu F_2(q^2)\biggr]u(p). \eqno(\nr)$$
The two well-known form factors are given by
$$\eqalignno{
F_1(q^2)&=
{\alpha\over2\pi} \int_0^1\rho \d\rho \int_0^1 \d x
\biggl[{1\over D(m^2,q^2)} \cr
&\null\times\bigl\{q^2[-(1-\rho)-\rho^2x^2]
+m^2[2(1-\rho)-\rho^2] \bigr\} \cr
&-(\hbox{same with }q^2=0) \biggr], &(\nr)\cr
F_2(q^2)&=
-{\alpha m^2\over\pi} \int_0^1\rho \d\rho \int_0^1 \d x
{\rho(1-\rho)\over D(m^2,q^2)}, &(\nr)\cr
D(m^2,q^2)&=\rho^2x(1-x)q^2-\rho^2 m^2 -(1-\rho)\lambda^2+\ieps &(\nr)}$$

For $-q^2\gg m^2$, one finds
\ref{\refKallen}
{G. K\"allen, in {\it Handbuch der Physik} vol. V/1 (Springer, 1958) 169}
$$\eqalignno{
F_1(q^2)&=
{\alpha\over4\pi}\biggl[
-\log{|q^2|\over m^2}
\biggl(\log{|q^2|\over\lambda^2} +\log{m^2\over\lambda^2}\biggr) \cr
&\phantom{{\alpha\over4\pi}\biggl[}
+2\log{m^2\over\lambda^2} +3\log{|q^2|\over m^2}
+{\pi^2\over3}-4 \biggr], &(\nr) \cr
F_2(q^2)&=
-{\alpha m^2\over\pi q^2} \log\biggl({|q^2|\over m^2}\biggr). &(\nr) }$$
\newappendixtext{C}{The $N_I$ and $N_{III}$-parts of diagram (a)}
We shall in this appendix show that the $N_I$ and $N_{III}$-parts
of diagram (a) behave like $1/s$.

\vskip8pt
\noindent C.1\ THE $N_I$-PART OF DIAGRAM $(a)$

Let
$$I_I(s,t)=\int_0^1\ldots\int_0^1 \d\alpha_1\ldots \d\alpha_7
\delta(1-\sum\alpha_i)\, {N_I\over \Lambda(\alpha)^3\, D(\alpha)},
\eqno(\nr)$$
with [cf.\ eq.~{\eqNone}]
$$N_I=8[5\Lambda(\alpha)-3\alpha_2(\alpha_4+\alpha_6+\alpha_7)].
\eqnref{\eqEtwo}$$
Since, according to the assumption $I_I(s,t)\sim1/s$,
we perform a Mellin transformation
[cf.\ eqs.~{\eqMellin} and {\eqdlower}]
$$\eqalignno{
\tilde I_I(\zeta)&=\int_0^\infty\d s\, s^{-\zeta}\, I(s,|t|=1) \cr
&={\Gamma(1-\zeta)\Gamma(\zeta)\over\Gamma(1)}\,
\int_0^1\ldots\int_0^1 \d\alpha_1\ldots \d\alpha_7
\delta(1-\sum\alpha_i) \,
{N_I\over \Lambda(\alpha)^3}\,
D_s^{-1+\zeta}\, d^{-\zeta} \cr
&\simeq{1\over\zeta}\,
\int_0^1\ldots\int_0^1 \d\alpha_1\ldots \d\alpha_7
\delta(1-\sum\alpha_i)\,
{N_I\over \Lambda(\alpha)^3}\,
(\alpha_2\alpha_4\alpha_5)^{-1+\zeta}. &(\nr)}$$
The dominant contribution is seen to come from
$\alpha_2\simeq\alpha_4\simeq\alpha_5\simeq0$.
However, we cannot evaluate the triple integral over $\alpha_2$,
$\alpha_4$ and $\alpha_5$ by putting $\alpha_2=\alpha_4=\alpha_5=0$
in $N_I/\Lambda^3$, since that procedure would
lead to a divergent integral in the remaining four variables.
Instead, we apply the Cheng-Wu theorem.
Approximating [cf.\ eqs.~{\eqEtwo}]
$$N_I\simeq40\Lambda(\alpha), \eqno(\nr)$$
and writing [cf.\ eq.~{\eqLambda}]
$$\eqalignno{
\Lambda(\alpha)
&=\alpha_2(\alpha_1+\alpha_3+\alpha_4+\alpha_5+\alpha_6+\alpha_7)
+(\alpha_1+\alpha_3+\alpha_5)(\alpha_4+\alpha_6+\alpha_7) \cr
&\equiv A\alpha_2+B, &(\nr)}$$
we obtain
$$\eqalignno{
\tilde I_I(\zeta)
&\simeq{40\over\zeta}\,
\int_0^\infty\d\alpha_2 \int_0^\infty\d\alpha_4 \int_0^\infty\d\alpha_5
\int_0^1\cdots \int_0^1\d\alpha_1 \d\alpha_3 \d\alpha_6 \d\alpha_7\,
\delta(1-\alpha_1-\alpha_3-\alpha_6-\alpha_7)\cr
&\times(\alpha_2\alpha_4\alpha_5)^{-1+\zeta}\,
(A\alpha_2+B)^{-2}.
\eqalref{\eqCsix}}$$
We first perform the $\alpha_2$-integration.
Invoking the Euler $\beta$ function [cf.\ eq.~{\eqBeta}] we have
$$\int_0^\infty\d\rho\, {\rho^{-1+\zeta}\over (A\rho+B)^{2-\nu}}
={\Gamma(\zeta)\Gamma(2-\nu-\zeta)\over\Gamma(2-\nu)}\,
A^{-\zeta}\, B^{-2+\nu+\zeta}. \eqnref{\eqEuler}$$
The leading term of the integral {\eqCsix} is thus, with $\nu=0$,
$$\eqalignno{
\tilde I_I(\zeta)
&\simeq{40\over\zeta^2}\,
\int_0^1\d\alpha_1 \int_0^1\d\alpha_3
\int_0^1\d\alpha_6 \int_0^1\d\alpha_7\,
\delta(1-\alpha_1-\alpha_3-\alpha_6-\alpha_7) &(\nr) \cr
&\times\int_0^\infty\d\alpha_4 \, \alpha_4^{-1+\zeta}
(\alpha_4+\alpha_6+\alpha_7)^{-2+\zeta}\,
\int_0^\infty\d\alpha_5 \, \alpha_5^{-1+\zeta}
(\alpha_1+\alpha_3+\alpha_5)^{-2+\zeta}.
}$$
Applying now eq.~{\eqEuler}, with $\nu=\zeta$, we get
$$\eqalignno{
\tilde I_I(\zeta)
&\simeq{40\over\zeta^4}\,
\int_0^1\d\alpha_1 \int_0^1\d\alpha_3
\int_0^1\d\alpha_6 \int_0^1\d\alpha_7\,
\delta(1-\alpha_1-\alpha_3-\alpha_6-\alpha_7)\cr
&\times (\alpha_1+\alpha_3)^{-2+2\zeta}\,
(\alpha_6+\alpha_7)^{-2+2\zeta}. &(\nr)}$$
With
$$\alpha_1+\alpha_3=\rho, \qquad \alpha_6+\alpha_7=\rho', \eqno(\nr)$$
we get
$$\eqalignno{
\tilde I_I(\zeta)
&\simeq{40\over\zeta^4}\,
\int_0^1\rho\d\rho \int_0^1\rho'\d\rho'\, \delta(1-\rho-\rho')
(\rho\rho')^{-2+2\zeta} \cr
&={40\over\zeta^4}\,
\int_0^1\d\rho [\rho(1-\rho)]^{-1+2\zeta} \simeq{40\over\zeta^5}. &(\nr)}$$
We conclude that the dominant contribution from the $N_I$-part
behaves like
$$F_{00}^{\rm (a)}(s,t)\bigg|_{N_I}
\sim{1\over s}\, \log^4\biggl({s\over|t|}\biggr). \eqno(\nr)$$
which we may neglect as $s\to\infty$.

\vskip8pt
\noindent C.2\ THE $N_{III}$-PART OF DIAGRAM $(a)$

Let
$$I_{III}(s,t)=
\int_0^1\ldots\int_0^1 \d\alpha_1\ldots \d\alpha_7\,
\delta(1-\sum\alpha_i)\,
{2N_{III}\over \Lambda(\alpha)^3\, D(\alpha)^3},
\eqno(\nr)$$
with $N_{III}$, $D(\alpha)$ and $\Lambda(\alpha)$ given
by eqs.~{\eqDalpha} and {\eqDs}--{\eqNthree}.
Since, according to our assumption $I_{III}(s,t)\sim1/s$ we perform
a Mellin transformation
[cf.\ eqs.\ {\eqMellin} and {\eqdlower}]
and substitute for $N_{III}$,
$$\eqalignno{
\tilde I_{III}(\zeta)&=\int_0^\infty\d s\, s^{-\zeta}\, I_{III}(s,|t|=1) \cr
&={2\Gamma(3-\zeta)\Gamma(\zeta)\over\Gamma(3)}\,
\int \d^7\alpha\, {N_{III}/s^2\over \Lambda(\alpha)^3}\,
D_s^{-3+\zeta}\, \calD(\alpha)^{-\zeta} \cr
&\simeq{2\over\zeta}\,
\int \d^7\alpha\, {\alpha_2^3\alpha_4^2\alpha_5^2
(\alpha_4+\alpha_6+\alpha_7)\over \Lambda(\alpha)^3}\,
(\alpha_2\alpha_4\alpha_5)^{-3+\zeta}\, \calD(\alpha)^{-\zeta} \cr
&\simeq{2\over\zeta}\,
\int \d^7\alpha\, {\alpha_4+\alpha_6+\alpha_7 \over \Lambda(\alpha)^3}\,
\alpha_2^\zeta(\alpha_4\alpha_5)^{-1+\zeta}\, \calD(\alpha)^{-\zeta}, &(\nr)}$$
where $\calD$ is given by eq.~{\eqdlower}.

To leading order, the integrations over $\alpha_4$ and $\alpha_5$
yield two factors of $1/\zeta$, whereas the remaining integrations
are non-singular.  Consequently, this Mellin transform shows that
$$I_{III}(s,t)\sim{1\over s}\,\log^2\biggl({s\over|t|}\biggr), \eqno(\nr)$$
which may be neglected as compared with the $1/t$ term of sect.~4.

\newappendixtext{D}{The remainder $R$}
In this appendix we evaluate the remainder $R$ of eqs.~{\eqRdef}.
We have
$$\eqalignno{
R(m^2,\lambda^2)&=\int_0^1\d \alpha\int_0^1{\d\rho\over\rho}
\biggl[{1\over 1+\rho}\,{1\over N} -{1\over N_0}\biggr] \cr
&+\int_0^1\d \alpha\int_1^\infty {\d \rho\over \rho}\, {1\over 1+\rho}\,
{1\over N} \cr
&=R_1+R_2, \eqalref{\eqGone}}$$
with
$$\eqalignno{
N&=\alpha(1-\alpha)+\rho^2 m^2+(1+\rho)\lambda^2, \cr
N_0&=N(\rho=0)=\alpha(1-\alpha)+\lambda^2, \eqalref{\eqNN}}$$

Here we rearrange the first part as
$$R_1=\int_0^1\d \alpha
\biggl[\int_0^1{\d\rho\over\rho}\biggl({1\over N}-{1\over N_0}\biggr)
-\int_0^1{\d\rho\over1+\rho}\, {1\over N}\biggr]. \eqno(\nr)$$
The integration over $\alpha$ is done in terms of the function $C_{20}$
of eq.~{\eqCtwozero},
$$R_1=-2\int_0^1{\d\rho\over\rho}\,
\log\biggl[\rho^2{m^2\over\lambda^2}+(1+\rho)\biggr]
+2\int_0^1{\d\rho\over 1+\rho}\,
\log[\rho^2m^2+(1+\rho)\lambda^2]. \eqno(\nr)$$
In the second integral the $\lambda^2$-term can be neglected.

Introducing
$$a_{1,2}={\lambda\over2m}\pm i\sqrt{1-{\lambda^2\over4m^2}}, \eqno(\nr)$$
we get
$$R_1=2\biggl[\Li2\biggl(-{m\over a_1\lambda}\biggr)
+\Li2\biggl(-{m\over a_2\lambda}\biggr)\biggr]
+2\log2\,\log(m^2)+4\,\Li2(-1), \eqno(\nr)$$
where
we have introduced the dilogarithmic function
{\refKolbig},
{\refDevoto}:
$$\eqalignno{
\Li2(z)
&=-\int_0^1{dt\over t}\,\log(1-zt) \cr
&=-\int_0^z{dt\over t}\,\log(1-t), \eqalref{\eqLi}}$$
with
$$\Li2(1)={\pi^2\over6}, \qquad
\Li2(-1)=-{\pi^2\over12}. \eqno(\nr)$$
Exploiting next
$$\Li2(-y^{-1})=-\Li2(-y)-\half\log^2(y)-{\pi^2\over6}, \eqno(\nr)$$
we arrive at
$$R_1=-2\log^2\biggl({\lambda\over m}\biggr)
+2\log2\,\log(m^2) -{\pi^2\over2}. \eqno(\nr)$$

In the second part of eq.~{\eqGone}, we can neglect the
$\lambda$-dependence.
We invert the variable, $\rho\to\rho^{-1}$,
$$R_2=\int_0^1{\d\rho\over 1+\rho}\int_0^1\d \alpha\,
{1\over \alpha(1-\alpha)+(m^2/\rho^2)}. \eqno(\nr)$$
The integration over $\alpha$ can be done with the result
$$R_2=\int_0^1{\d\rho\over 1+\rho}\,
{1\over\sqrt{{1\over4}+{m^2\over\rho^2}}}\,
\biggl[2\log\biggl({1\over2}+\sqrt{{1\over4}+{m^2\over\rho^2}}\biggr)
-\log\biggl({m^2\over\rho^2}\biggr)\biggr]. \eqno(\nr)$$
The first term in the parenthesis yields a vanishing contribution
in the limit $m\to0$.
In the second term we can let $m\to0$ in the square root factor,
giving
$$\eqalignno{
R_2&=-2\int_0^1{\d\rho\over 1+\rho}
\log\biggl({m^2\over\rho^2}\biggr) \cr
&=-2\log2\,\log(m^2)+4\,\Li2(-1). &(\nr)}$$

Adding the two contributions, we get
$$R(m^2,\lambda^2)
=-{1\over2}\log^2\biggl({\lambda^2\over m^2}\biggr)-{5\pi^2\over6}.
\eqno(\nr)$$

The integral encountered in eq.~{\eqItwotilde} is easily related to $R$.
A transformation of variable, $x=1/(1+\rho)$ gives
$$\eqalignno{
&\int_0^1\d z\int_0^1\d x\, {1\over 1-x}
\biggl[{x^2\over a_0} -{1\over a_{00}}\biggr] \cr
&=\int_0^1\d z \int_0^\infty{\d\rho\over \rho(1+\rho)}
\biggl[{1\over N(z)}-{1\over N_0(z)}\biggr] \cr
&=R-
\int_0^1\d z{1\over N_0(z)}
\int_1^\infty{\d\rho\over \rho(1+\rho)}
+\int_0^1\d z{1\over N_0(z)}
\int_0^1{\d\rho\over 1+\rho} =R. &(\nr)}$$
Here, $a_0$ and $a_{00}$ are defined in eq.~{\eqaa}
and $N(z)$ and $N_0(z)$ are obtained from the expressions
in eq.~{\eqNN} by substituting $z$ for $\alpha$.

\newappendixtext{E}{The contributions $F_{00}^{\rm(b2)}$
and $F_{00}^{\rm(b3)}$}
\vskip8pt
\noindent E.1\ THE TERM $F_{00}^{\rm(b2)}$ OF DIAGRAM $(b)$

For the second term of eqs.~{\eqGamma} and {\eqFonethree}, we get,
in analogy with eq.~{\eqFzerozeroone}
$$\eqalignno{
F_{00}^{\rm(b2)}(s,t)&=2\int_0^1 \rho \d\rho \int_0^1 \d x
\int_0^1\ldots\int_0^1 \d\alpha_1\ldots \d\alpha_4 \,
\delta(1-\sum\alpha_i){n_2\over \Lambda(\beta)^2} \cr
&\null\times\biggl[{N_{II}^{\hbox{\eightrm box}}(\beta)\over D(\beta)^2}
+{N_{I}^{\hbox{\eightrm box}}(\beta)\over 2D(\beta)} \biggr], &(\nr) \cr
&\equiv I_{II}(s,t)+I_I(s,t), \eqalref{\eqFbtwo} }$$
with
$$\eqqqalignno{
\beta_1&=\alpha_1+\alpha_3 r_1, & \beta_2&=\alpha_2, \cr
\beta_3&=\alpha_3 r_2,          & \beta_4&=\alpha_4, &(\nr)}$$
and
$$\eqalignno{
\Lambda(\beta)&=\beta_1+\beta_2+\beta_3+\beta_4 \cr
&=\alpha_1+\alpha_2+\alpha_3(r_1+r_2)+\alpha_4, &(\nr) \cr
D(\beta)&=D_{\hbox{\eightrm box}}(\beta)
-\alpha_3\Lambda(\beta) M^2 \cr
&=\beta_1\beta_2s+\beta_3\beta_4t
-[(\beta_1+\beta_2)^2+\alpha_3\Lambda(\beta)\rho^2]m^2 \cr
&-[(\beta_3+\beta_4)+\alpha_3(1-\rho-r_2)]\Lambda(\beta)\lambda^2
+\ieps. \eqalref{\eqdElower} }$$

We perform a Mellin transform on the first term of eq.~{\eqFbtwo},
$$\eqalignno{
\tilde I_{II}(\zeta)&=\int_0^\infty \d s\, s^{-\zeta-1}\, I_{II}(s,|t|=1) \cr
&={2\Gamma(1-\zeta)\Gamma(1+\zeta)\over \Gamma(2)}\,
\int_0^1\rho\d\rho \int_0^1\d x
\int_0^1\ldots\int_0^1 \d\alpha_1\ldots\d\alpha_4 \,
\delta(1-\sum\alpha_i) \cr
&\times {n_2\over\Lambda(\beta)^2}
\biggl({N_{II}^{\hbox{\eightrm box}}(\beta)\over s}\biggr)
D_s^{-1+\zeta}\, d^{-1-\zeta}, &(\nr)}$$
where
$$D_s=\beta_1\beta_2=[\alpha_1+\alpha_3\rho(1-\rho)x]\alpha_2, \eqno(\nr)$$
and $d$ is given by {\eqdElower} as
$$\eqalignno{
-d
&=\beta_3\beta_4+[(\beta_1+\beta_2)^2+\Lambda(\beta)\alpha_3\rho^2]m^2 \cr
&+\{(\beta_3+\beta_4)+\alpha_3[1-\rho-\rho^2x(1-x)]\}\Lambda(\beta)\lambda^2.
&(\nr)}$$
It is now seen that the integration over $\alpha_2$ yields a factor $1/\zeta$,
but that no further singularity is obtained from the remaining integrations.
Thus, this is a contribution to $A_1$ of eq.~{\eqtwotwo}.
It is cancelled by the corresponding contribution from the crossed diagram,
(${\rm b}'$).

The second term of eq.~{\eqFbtwo} behaves at large $s$ as $1/s$ and may be
neglected.

\vskip8pt
\noindent E.2\ THE TERM $F_{00}^{\rm(b3)}$ OF DIAGRAM $(b)$

The amplitude corresponding to the $F_{00}^{\rm(b3)}(s,t)$ of eqs.~{\eqGamma}
and {\eqFonethree} is given by
$$\eqalignno{
{\cal M}_{\rm(b3)}
&=(-ie)^4(M^\gamma)_{\mu\nu}(M_q)^{\mu\nu} \cr
&\equiv{i\alpha^3\over4\pi}\, F_{00}^{\rm(b3)}(s,t)
[\bar u(p_1')\gamma^\mu u(p_1)]\;
[\bar v(p_2)\gamma_\mu v(p_2')] &(\nr)}$$
with
$$\eqalignno{
(M^\gamma)_{\mu\nu}
&=[\bar u(p_1')\gamma^\beta u(p_1)]\;
[\bar v(p_2)\gamma_\mu\gamma_\nu\gamma_\beta v(p_2')], &(\nr) \cr
(M_q)^{\mu\nu}
&={\alpha\over2\pi}\int_0^1\rho \d\rho \int_0^1 \d x \, f(\rho,x) \cr
&\null\times \int{\d^4q\over(2\pi)^4}\,
{q^\mu(-p_2-q)^\nu \over D((p_1-q)^2, q^2) [(p_2+q)^2-m^2+\ieps]} \cr
&\null\times {1\over q^2-\lambda^2+\ieps} \,
{1\over (p_1-p_1'-q)^2-\lambda^2+\ieps}, &(\nr)}$$
and
$$\eqalignno{
D((p_1-q)^2, q^2)&=
r_1[(p_1-q)^2-m^2] +r_2[q^2-\lambda^2]-M^2+\ieps, &(\nr) \cr
f(\rho,x)&=-2\rho x(1-\rho x). &(\nr)}$$
with $r_1$ and $r_2$ defined by eq.~{\eqmdef} and $M^2$ by eq.~{\eqMtwo}.

The integration over momenta is performed as in sect.~3.1 and gives
$$\eqalignno{
(M_q)^{\mu\nu}
&={\alpha\over2\pi}\int_0^1\rho \d\rho \int_0^1 \d x \, f(\rho,x) \cr
&\null\times {i\over16\pi^2}
\int_0^1 \d\alpha_1\cdots \d\alpha_4\, \delta(1-\sum\alpha_i)\,
{1\over\Lambda(\gamma)^2} \cr
&\null\times\biggl[{1\over D(\gamma)^2}\, k_2^\nu k_3^\mu
-{1\over 2D(\gamma)}\, g^{\mu\nu}\biggr], &(\nr)}$$
with
$$\eqqqalignno{
\gamma_1&=\alpha_1r_1,           & \gamma_2&=\alpha_2, \cr
\gamma_3&=\alpha_3+\alpha_1r_2,  & \gamma_4&=\alpha_4, &(\nr) }$$
$$\eqalignno{
D(\gamma)&=D_{\hbox{\eightrm box}}(\gamma)
-\alpha_1\Lambda(\gamma)M^2, \cr
\Lambda(\gamma)&=\gamma_1+\gamma_2+\gamma_3+\gamma_4, \cr
M^2&=\rho^2m^2 +(1-\rho)\lambda^2-r_2\lambda^2, &(\nr)}$$
and
$$\eqalignno{
k_2&=-\gamma_1p_1-(\gamma_1+\gamma_3+\gamma_4)p_2-\gamma_4(p_1-p_1'), \cr
k_3&=\gamma_1p_1-\gamma_2p_2+\gamma_4(p_1-p_1'). &(\nr)}$$
We find
$$\eqalignno{
F_{00}^{\rm(b3)}(s,t)&=
\int_0^1\rho \d\rho\int_0^1 \d x \, f(\rho,x)
\int \d\alpha_1\cdots \d\alpha_4 \, \delta(1-\sum\alpha_i) \cr
&\null\times{1\over\Lambda(\gamma)^2} \,
\biggl[{N_{II}\over D(\gamma)^2} +{N_I\over 2D(\gamma)}\biggr] \cr
&\equiv I_{II}(s,t)+I_{I}(s,t), &(\nr) \cr
N_{II}&=-2s\gamma_1(\gamma_1+\gamma_3+\gamma_4), \cr
N_I&=-8. &(\nr)}$$

A Mellin transform,
$$\tilde I_{II}(\zeta)=
\int_0^\infty \d s\,s^{-\zeta-1}\, I_{II}(s,|t|=1), \eqno(\nr)$$
shows that the most singular term is proportional to $1/\zeta$,
implying $I_{II}(s,t)\sim (1/t)$. Since there is no factor
$\log s$,
this contribution is cancelled against the corresponding one from the
crossed diagram.

The last term, $I_{I}(s,t)$ is similarly found to behave like
$(1/s)\log^3(s)$ and may therefore be ignored at large values of $s$.

\vfill\eject\immediate\closeout\rfile
\baselineskip=14pt\centerline{{\bf References}}\bigskip{\frenchspacing%
\catcode`\@=11\escapechar=` %
\input refs.tmp\vfill\eject}\nonfrenchspacing
\centerline{\bf Figure captions}

\vskip 15pt
\def\fig#1#2{\hangindent=.65truein \noindent \hbox to .65truein{Fig.\ #1.
\hfil}#2\vskip 2pt}

\fig1{The eight Feynman diagrams considered.}

\fig2{Choice of Feynman parameters for the box diagram.}

\fig3{Choice of Feynman parameters for the uncrossed $(a)$ diagram
of fig.~1.}

\fig4{Vertex correction to the box diagram, corresponding to
diagram $(b)$ of fig.~1.}

\fig5{The electron self energy.}

\fig6{The vertex function.}

\bye